\documentclass[conference]{IEEEtran}
\IEEEoverridecommandlockouts
\usepackage{cite}
\usepackage{amsmath,amssymb,amsfonts}
\usepackage{graphicx}
\usepackage{textcomp}
\usepackage{xcolor}
\usepackage[hyphens]{url}

\usepackage{xspace}
\newcommand{\sysname}[0]{{\textsc{NasZip}}\xspace}

\usepackage{amsmath,amsfonts,bm}

\def\eqref#1{equation~\ref{#1}}
\def\floor#1{\lfloor #1 \rfloor}
\def\1{\bm{1}}

\def\vb{{\bm{b}}}
\def\vc{{\bm{c}}}

\def\vp{{\bm{p}}}
\def\vq{{\bm{q}}}

\def\vs{{\bm{s}}}

\def\vv{{\bm{v}}}

\def\vx{{\bm{x}}}

\def\mP{{\bm{P}}}

\DeclareMathAlphabet{\mathsfit}{\encodingdefault}{\sfdefault}{m}{sl}
\SetMathAlphabet{\mathsfit}{bold}{\encodingdefault}{\sfdefault}{bx}{n}

\def\sC{{\mathbb{C}}}

\def\sP{{\mathbb{P}}}

\newcommand{\normltwo}{L^2}

\usepackage{annotate-equations}

\usepackage{soul}
\soulregister\cref7
\soulregister\cite7
\soulregister\textrm7
\soulregister{\aka}{7}
\soulregister{\vs}{7}
\soulregister{\wrt}{7}
\soulregister{\sysname}{7}
\soulregister\ballnumberA7

\usepackage{multicol}
\usepackage{multirow}
\usepackage{makecell}
\usepackage{booktabs}
\usepackage{enumitem}

\definecolor{note-color}{rgb}{1,0,0}
\definecolor{hl_pink}{RGB}{255,228,225}

\makeatletter
\DeclareRobustCommand\onedot{\futurelet\@let@token\@onedot}
\def\@onedot{\ifx\@let@token.\else.\null\fi\xspace}

\def\eg{\emph{e.g}\onedot} 
\def\ie{\emph{i.e}\onedot} 
 
 \def\vs{\emph{vs}\onedot}
\def\wrt{w.r.t\onedot} 

\def\aka{\emph{aka}\onedot} 
\makeatother

\usepackage{tikz}
\newcommand{\ballnumberA}[1]{\tikz[baseline=(myanchor.base)] \node[circle,fill=.,inner sep=1pt] (myanchor) {\color{-.}\bfseries\footnotesize #1};}
\newcommand{\ballnumberB}[1]{\tikz[baseline=(myanchor.base)] \node[circle,draw=black,fill=none,inner sep=1pt] (myanchor) {\color{black}\bfseries\footnotesize #1};}

\usepackage{algorithm}
\usepackage{algorithmicx}
\usepackage{algpseudocode}
\usepackage{subcaption}
\usepackage{makecell} % put this in the preamble

\newcommand{\design}[1]{\textit{#1}}

\def\BibTeX{{\rm B\kern-.05em{\sc i\kern-.025em b}\kern-.08em
    T\kern-.1667em\lower.7ex\hbox{E}\kern-.125emX}}
\usepackage{hyperref}
\hypersetup{
    colorlinks=true,
    citecolor=blue,
    linkcolor=blue,
    filecolor=magenta,      
    urlcolor=cyan,
    pdfpagemode=FullScreen,
}

\usepackage[capitalise]{cleveref}

\makeatletter
\newcommand{\linebreakand}{%
  \end{@IEEEauthorhalign}
  \hfill\mbox{}\par
  \mbox{}\hfill\begin{@IEEEauthorhalign}
}
\makeatother

\begin{document}

% Ensure letter paper
\pdfpagewidth=8.5in
\pdfpageheight=11in

%%%%%%%%%%%---SETME-----%%%%%%%%%%%%%
\newcommand{\iscasubmissionnumber}{122}
%%%%%%%%%%%%%%%%%%%%%%%%%%%%%%%%%%%%

% \pagenumbering{arabic}

%%%%%%%%%%%---SETME-----%%%%%%%%%%%%%
\title{NasZip: Software and Hardware Co-Design to Accelerate Approximate Nearest Neighbor Search with DIMM-Based Near-Data Processing}

\author{
\IEEEauthorblockN{
Cheng Zou\IEEEauthorrefmark{1}\IEEEauthorrefmark{6},
Shuo Yang\IEEEauthorrefmark{4}\IEEEauthorrefmark{1}\IEEEauthorrefmark{6},
Chen Nie\IEEEauthorrefmark{1}\IEEEauthorrefmark{2},
Yu Zou\IEEEauthorrefmark{3},
Yu He\IEEEauthorrefmark{5},
Chao Jiang\IEEEauthorrefmark{5},\\
Limin Xiao\IEEEauthorrefmark{5},
Weifeng Zhang\IEEEauthorrefmark{5},
Zhezhi He\IEEEauthorrefmark{1}\IEEEauthorrefmark{2}\IEEEauthorrefmark{7}
},
\IEEEauthorblockA{\IEEEauthorrefmark{1}Intelligent Computing Research Group, School of Computer Science, Shanghai Jiao Tong University, Shanghai, CN
}
\IEEEauthorblockA{\IEEEauthorrefmark{2}Shanghai AI Laboratory, Shanghai, CN;
\IEEEauthorrefmark{3}Institute of Information Engineering, Chinese Academy of Sciences, Beijing, CN
}
\IEEEauthorblockA{\IEEEauthorrefmark{4}School of Integrated Circuits, Shanghai Jiao Tong University, Shanghai, CN;
\IEEEauthorrefmark{5}Lenovo Research, Beijing, CN
}
\IEEEauthorblockA{
Email: chenchen\_zou@sjtu.edu.cn, yangshuo1230@sjtu.edu.cn, zhezhi.he@sjtu.edu.cn
}
\IEEEauthorblockA{
\IEEEauthorrefmark{6} Equal contribution, \IEEEauthorrefmark{7}Corresponding author
}
}

\maketitle

%%%%%% -- PAPER CONTENT STARTS-- %%%%%%%%

\begin{abstract}

As large language models (LLMs) continue to advance, retrieval-augmented generation (RAG) has become the key mechanism for expanding model knowledge and reducing hallucinations.
Central to RAG is approximate nearest neighbor search (ANNS), which retrieves database vectors most similar to a given query.
However, distance calculation over high-dimensional vectors is inherently memory-bound, causing retrieval performance to be constrained by I/O bandwidth on mainstream platforms such as CPUs and GPUs.
Although many prior early exiting (EE) techniques attempt to reduce memory accesses by only computing partial dimensions, the partial distance converges too slowly to the EE threshold, which ultimately limits their performance gains.
To address these challenges, we propose \sysname, a hardware-software co-designed framework that integrates near-data processing (NDP) with a novel feature-level early exiting guided by statistics-based principal component analysis (PCA).
Instead of relying solely on partial distances, \sysname incorporates estimation and correction parameters to approximate full-dimensional distances accurately, enabling \textit{earlier exiting without compromising accuracy}.
We further introduce a bit-level NDP-aware dynamic-float scheme that significantly reduces memory access for vector data.
On the hardware side, we develop a data-aware neighbor list mapping strategy that reduces neighbor-retrieval latency and inter-channel communication overhead, complemented by a dedicated cache that exploits data locality and enhances prefetch efficiency.
With these co-optimized techniques, \sysname delivers speedups of up to 8.4$\times$/1.4$\times$ over CPU baseline and state-of-the-art GPU implementation at equal accuracy.  
Relative to the state-of-the-art NDP ANNS accelerator ANSMET, \sysname achieves 1.69$\times$ performance improvement.

\end{abstract}

\section{Introduction}

Large language models (LLMs) have demonstrated remarkable capabilities across diverse tasks~\cite{naveed2023comprehensive}.
To enhance their factual grounding and adaptability, retrieval‑augmented generation (RAG) has emerged by enabling LLMs to query and integrate external knowledge during inference \cite{lewis2020retrieval}.
At the core of RAG lies approximate nearest neighbor search (ANNS), which retrieves semantically relevant content from large-scale vector databases \cite{peng2023efficient}.
ANNS typically consists of two stages: \textit{index construction}, where the corpus vectors are organized into a searchable structure, and \textit{query search}, where relevant vectors are identified based on embedding proximity.

There are various kinds of index construction methods, including tree-based \cite{ben1975tree,muja2014tree,balltree}, hash-based \cite{gan2012C2LSH,das2011lsh,datar2004LSH}, cluster-based \cite{2011ivf}, and graph-based \cite{dong2011graphknn, fu2019nsg, fu2022ssg, malkov2018hnsw, ootomo2024cagra} approaches.
Among them, graph-based ANNS (GANNS) offers superior performance with high accuracy and low latency \cite{gao2023anns}, and thus forms the focus of this work.

In contrast to the one-time index construction, the query search is executed repeatedly during LLM inference and therefore determines overall system performance.
The distance computations between the query vector and candidate vectors exhibit low arithmetic intensity, making performance fundamentally limited by the data access bandwidth.
Prior studies \cite{li2025ansmet,wang2024ndsearch} introduce early exiting (EE) techniques that compute distances over partial vector dimensions and terminate once the accumulated partial distance exceeds a predefined threshold.
However, partial distances still require a relatively large number of dimensions to converge to the threshold, leading to conservative performance gains ($\sim$ 20\%).

As a countermeasure, we propose a \textit{feature-level EE with statistics-based PCA} (FEE-sPCA), which estimates full vector distances from partial computing results to enable earlier exiting. 
We propose a bit-level compression scheme using \textit{dynamic floating-point} (Dfloat) to further reduce data accesses.

Meanwhile, due to the memory-bound nature of the ANNS, NDP devices are increasingly adopted to improve performance by utilizing the high internal memory bandwidth \cite{li2025ansmet,Junhyeok2023CXL-ANNS,quinn2025drex}.
ANSMET \cite{li2025ansmet} and CXL-ANNS \cite{Junhyeok2023CXL-ANNS} place customized acceleration logic near DRAM chips, while DReX \cite{quinn2025drex} deploys PIM (processing-in-memory) units to accelerate the filter stage in the retrieval.
However, their effectiveness remains limited.
We break down the latency for a naive NDP, and observe that existing designs \cite{li2025ansmet} cannot fully exploit NDP's internal bandwidth.
For example, the CPU-side ANNS index processing incurs a heavy overhead and causes heavy cross-channel data communications (occupying over 50\% ANNS latency), because the CPU is not aware of low-level vector data mapping on NDPs.
This becomes a new performance bottleneck even though the distance calculation is optimized via early exiting.

To improve upon this, we introduce a \textit{data-aware index mapping} (DaM) and offload neighbor lookup operations to NDP.
DaM ensures that each node's index and vector data reside within the same channel, minimizing cross-channel communications and fully exploiting NDP's internal parallelism.
Additionally, we find that similar or repeated queries show locality when accessing graph nodes, \ie, frequent accesses to neighbor list entries and vector data entries. 
Therefore, we further design a local neighbor cache, consisting of a \textit{local neighbor cache for table} (LNC-T) and a \textit{local neighbor cache for data} (LNC-D), to exploit the locality of ANNS queries.

In summary, we propose \sysname, a software-hardware co-optimization to efficiently accelerate ANNS on DIMM-based NDP devices. More specifically, our contributions are:
\begin{itemize}[leftmargin=*,label={$\triangleright$}]
\item On the algorithm level, we propose a novel two-fold optimization (\emph{VD-Zip}) consisting of a feature-level early exiting algorithm (FEE-sPCA) and a bit-level dynamic floating-point representation (Dfloat) to reduce data accesses and computations while maintaining the recall rate. 

\item On the hardware level, we propose several dedicated architectural components to accelerate GANNS on NDPs. 
Data-aware neighbor list mapping (DaM) is proposed to offload neighbor list lookup from CPUs to NDPs and reduce communication between DRAM channels. 
A local neighbor cache (LNC) is proposed to exploit locality across ANNS queries and enable prefetching.
\end{itemize}

Evaluated on six datasets at the same recall, \sysname achieves up to 8.4$\times$ and 1.4$\times$ speedup over the CPU baseline \cite{guo2020scann} and the GPU implementation CAGRA \cite{ootomo2024cagra} on NVIDIA A100, respectively.
\sysname also achieves a 1.69$\times$ speedup over the SOTA NDP accelerator ANSMET \cite{li2025ansmet}.

\section{Background}

\subsection{Approximate Nearest Neighbor Search}
\label{sec:anns_bg}

\subsubsection{Basics} Nearest neighbor search retrieves vectors $\{\vp_i\}$ closest to a query $\vq$ from a vector database (VecDB) $\mP$ of size $n$, each with $d$ dimensions:
\begin{equation}
\centering
\{\vp_i\}=\arg \min_{\vp_i \in \mP}  || \vp_i - \vq ||_2,\quad \mP \in \mathbb{R}^{n\times d}
\label{eqt:nn_def}
\end{equation}
where $||\cdot||_2$ calculates the  $\normltwo$ norm (or inner product).
Retrieving $k$ closest vectors is known as \emph{k-nearest-neighbors} (kNN) search, with $\mathcal{O}(nd)$ complexity.
To accelerate the searching process, \textit{approximate nearest-neighbor search} (ANNS) is proposed to return approximately closest vectors
with sub-linear time complexity as well as minor accuracy loss.

ANNS methods can be categorized as \emph{hashing-based}, \emph{tree-based} \cite{datar2004LSH, gan2012C2LSH, huang2015QALSH}, \emph{graph-based} \cite{malkov2014nsw, malkov2018hnsw, fu2019nsg, fu2022ssg} and \emph{quantization-based} \cite{pq2011, ge2014opq, gao2024rabitq}.
The graph-based approach is now widely adopted in commercial databases \cite{2021milvus, 2022manu, pinecone2025} and advanced RAG systems \cite{he2025hnswRAG}, as it can deliver orders-of-magnitude speedups over quantization-based and tree-based methods while maintaining high recall \cite{2020annbench}.
Despite high throughput, ANNS methods remain difficult to deploy.
Under extremely high accuracy requirements \cite{quinn2025iks}, their advantage over kNN can diminish, motivating more robust ANNS designs that sustain throughput without sacrificing accuracy.

\begin{figure}[t]
    \centering
    \includegraphics[width=\linewidth]{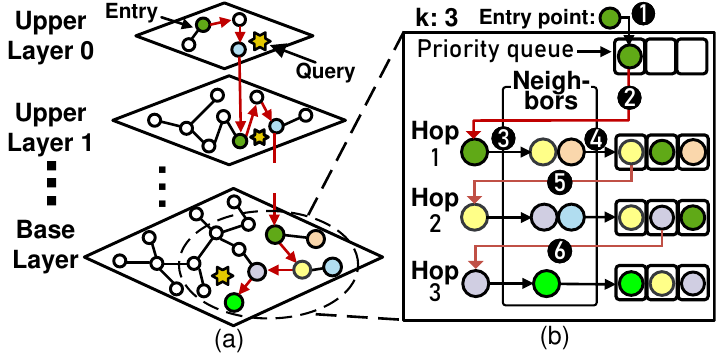}
    \caption{\textbf{An example multi-layer graph structure and breadth-first search (BFS) searching process for HNSW.}}
    \label{fig:hnsw}
\end{figure}

\subsubsection{Graph-based ANNS (GANNS)}
GANNS represents database vectors as graph nodes, with edges connecting similar nodes to enable efficient traversal toward target vectors in fewer hops.
Representative GANNS implementations include hierarchical navigable small worlds (HNSW) \cite{malkov2018hnsw} on CPUs and CAGRA \cite{ootomo2024cagra} on GPUs.
HNSW uses a multi-layer graph where the base layer contains all vectors, while upper layers contain fewer nodes and provide longer-range links for coarse-to-fine search.
In each traversal, the nearest node found at one layer serves as the entry point to the next lower layer.
CAGRA uses a single-layer graph for GPU efficiency, but its graph structure can be converted into the multi-layer form of HNSW \cite{ootomo2024cagra,malkov2018hnsw}.
\textit{We focus on HNSW in this work for generality.}

\subsubsection{Details of HNSW}
HNSW is illustrated in \cref{fig:hnsw}, with its graph structure in (a) and an example search procedure in (b).
The algorithm consists of a one-time index construction phase and a repeated search phase.
In this work, we focus on the search phase, as it dominates execution time.
Details of the hierarchical graph construction can be found in \cite{malkov2018hnsw}.

\cref{fig:hnsw} shows an example where a candidate priority queue keeps the top-$k$ nearest points found so far ($k=3$).
HNSW starts from the upper layer 0 and each layer's searching is a breadth-first search (BFS) process.
In each iteration, the unvisited closest point is selected from the candidate priority queue as the starting point for the next iteration.
The process is exemplified in \cref{fig:hnsw}b.
In \ballnumberA{1}, the entry point is added to the priority queue and used as the starting point for the first hop in \ballnumberA{2}.
Then, we access its neighbor list (\ballnumberA{3}), calculate their distances to the query, and insert them into the priority queue (\ballnumberA{4}).
Next, in \ballnumberA{5}, the closest point (yellow) from the priority queue is selected as the second hop's starting point, and we repeat \ballnumberA{3}\ballnumberA{4}.
The blue point is not added because it is not closer than the farthest green one in the queue.
In \ballnumberA{6}, the unvisited purple point is chosen as the third hop's starting point.
HNSW is controlled by two parameters, $efSearch$ and $M$.
$efSearch$ is the size of the candidate queue controlling the search scope during the online search stage.
$M$ is the maximum connections per node, controlling the graph density during the offline index construction stage.

\subsubsection{Evaluation metric}
The accuracy of ANNS is evaluated by the percentage of vectors $\sP'$, which are correctly identified by ANNS \wrt\  the ground truth (\ie, kNN result $\sP$).
It is denoted as $\textrm{recall}@k= |\sP' \cap \sP|/|\sP|$ under a top-$k$ search.

\subsection{Feature-Level Early Exiting}
\label{sec:fee_bg}

Before describing the feature-level early exiting (FEE) \cite{li2025ansmet,gao2023adsampling,gao2024rabitq}, we first define the terms in \cref{tab:notion}.
Given two $D$-dimensional vectors, the full distance ($d_{\textrm{all}}$) denotes the exact distance computed over all $D$ dimensions, while the partial distance ($d_{\textrm{part}}^{k}$) denotes the distance computed over only the first $k$ dimensions, where $k<D$. When $k=D$, the two are equal; otherwise, $d_{\textrm{part}}^{k}<d_{\textrm{all}}$ (L2).
As discussed in \cref{sec:anns_bg}, during each BFS step, a neighbor $\vx$ is added to the candidate queue only if its full distance to the query, $d_{\textrm{all}}(\vx,\vq)$, is smaller than the current farthest distance in the queue, denoted as $\mathit{threshold}$.
Otherwise, the computation is \textbf{wasted}.
FEE reduces this waste by terminating the $D$-dimensional distance computation early once the partial distance $d_{\textrm{part}}^{k}(\vx,\vq)$ exceeds $\mathit{threshold}$ after only $k$ dimensions.
We further analyze limitations of the existing approaches (\cref{subsec:moti_ndp_dist_cal}) and then present our improved design (\cref{sec:FEE}).

\begin{table}[t]
\caption{\textbf{Notations used in this work.}}
\label{tab:notion}
\renewcommand{\arraystretch}{1.1}
\footnotesize
\begin{tabular}
{>{\centering\arraybackslash}m{1.3cm}>
{\raggedright\arraybackslash}m{6.8cm}}
\hline
\textbf{Notion} & \multicolumn{1}{c}{\textbf{Description}} \\
\hline
$d_\textrm{part}^k(\vx,\vq)$ &
The partial distance of the first $k$ dimensions between vector $\vx$ and query vector $\vq$. \\
\hline
$d_\textrm{est}^k(\vx,\vq)$ &
The estimated distance of all dimensions between vector $\vx$ and query vector $\vq$. Estimation is based on $d_\textrm{part}^k(\vx,\vq)$. \\
\hline
$d_\textrm{all}(\vx,\vq)$ &
The real distance of all dimensions between vector $\vx$ and query vector $\vq$.  \\
\hline
$D$ &
The number of dimensions of vectors in the database.  \\
\hline
$threshold$ &
The distance between the query and the farthest vector in the candidate queue.  \\
\hline
\multicolumn{2}{@{}l}{\footnotesize *The distance noted here uses L2 norm or inner product distance.} \\
\end{tabular}
\end{table}

\subsection{Near-Data Processing (NDP)}
\label{sec:ndp}

\begin{figure}[t]
    \centering
    \includegraphics[width=\linewidth]{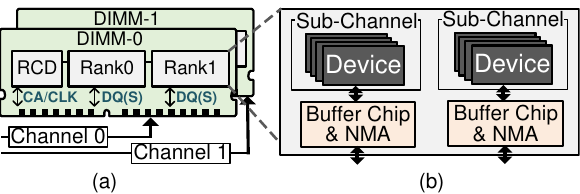}
    \caption{\textbf{Example illustration of DIMM-NDP.} (a) Two DIMMs are connected to two channels respectively. Each has one RCD chip, and several ranks. (b) Each rank has two sub-channels. Each sub-channel has four DRAM chips (device). The NMA is placed and packaged together with Buffer Chip of DIMM.}
    \label{fig:dimm}
\end{figure}

NDP \cite{mutlu2022modern} is widely explored to mitigate the ``memory wall'' by placing dedicated accelerators near memory, improving effective memory bandwidth by up to two orders of magnitude \cite{Gómez-Luna2022Benchmarking_a_new_paradigm, Lee20221ynm, Lee2021Industrial_product}.
Among existing paradigms, the dual in-line memory module (DIMM) based NDP is particularly attractive due to its large memory capacity, scalability, and technical maturity \cite{hass2022ndp}.
As shown in \cref{fig:dimm}a, a DIMM-based NDP system consists of DIMMs containing one register clock driver (RCD), several ranks, and multiple data buffers (DBs).
The RCD maintains control/address signal (CA/CLK) integrity, the DBs maintain data signal (DQ/DQS) integrity, and the ranks contain the high-density DRAM chips for data storage.

In this work, we target a DDR5-based DIMM-NDP architecture.
As shown in \cref{fig:dimm}b, each rank contains 8 DRAM devices organized into 2 sub-channels, each with 4 DRAM devices operating synchronously on the same offset.
Cross-channel communication is costly \cite{xie2025unindp,hass2022ndp,Tian2025ndpbridge} because sub-channels have no direct communication path and must exchange data through the processor.
To exploit sub-channel parallelism, a near-memory accelerator (NMA) is placed in each sub-channel.
By making only minor changes to the DB chips and interface, the design preserves host compatibility and reuses the existing processor DDR controller for practical programmability and software integration \cite{Tian2025ndpbridge,huangfu2019medal}.
In \sysname, the NMA logic is integrated into the DB chip, similar to \cite{li2025ansmet}, without modifying standard DRAM chips.

\section{Motivation}

\begin{figure}[t]
    \centering
    \includegraphics[width=\linewidth]{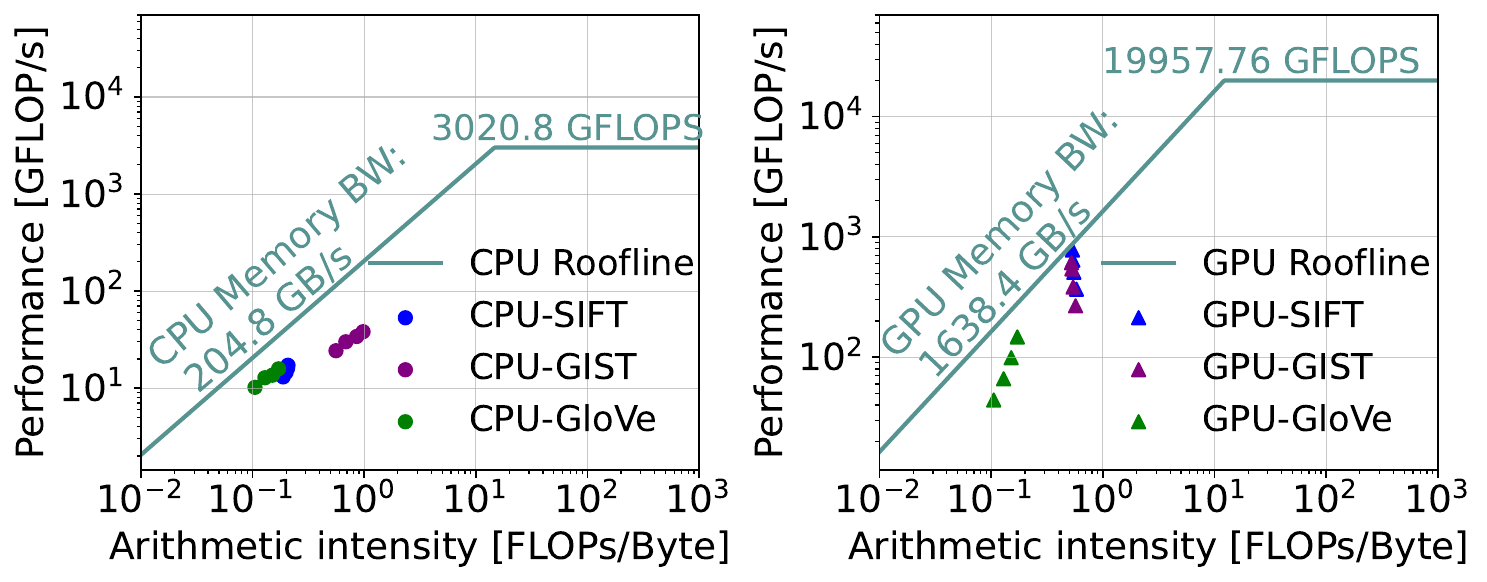}
    \caption{\textbf{The roofline model of ANNS implementations}
    on various datasets with CPU (left) and GPU (right).
    Testing configurations are given in \cref{sec:evaluation_methodology}.}
    \label{fig:ANNS_roofline}
\end{figure}

\subsection{Memory-Bound Nature of ANNS}

\cref{fig:ANNS_roofline} uses the roofline model \cite{williams2009roofline} to analyze ANNS, using different VecDB datasets (SIFT, GIST \cite{lowe2004sift}, and GloVe \cite{pennington2014glove}).
HNSW on CPU and CAGRA on GPU are both memory-bound, motivating us to leverage high aggregated internal memory bandwidth for higher performance.
Recent SRAM-based processing-in/near-memory designs leverage high-speed on-chip SRAM arrays for computation~\cite{MAICC,Nie2024VSPIM,polymorpic}.
However, their limited capacity and high per-bit cost make them unsuitable for storing large-scale vector databases.
Therefore, we leverage DIMM-based near-data processing to combine large DRAM capacity with high internal memory bandwidth.

\begin{figure}[t]
\begin{subfigure}[h]{0.55\linewidth}
    \centering
    \includegraphics[width=\linewidth]{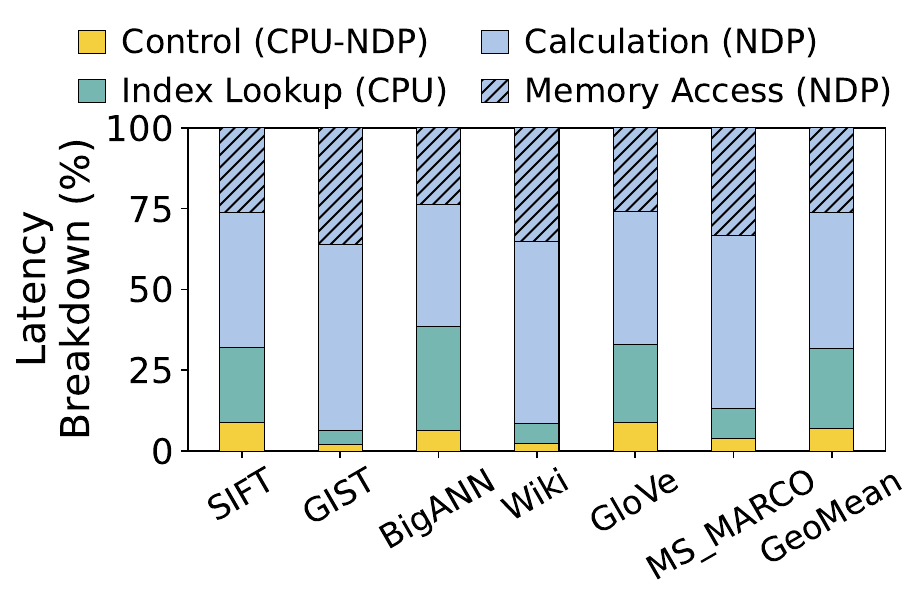}
    \vspace{-2em}
    \caption{ANNS latency breakdown}
    \label{fig:motivation_breakdown}
\end{subfigure}
    \hfill
\begin{subfigure}[h]{0.44\linewidth}
    \centering
    \includegraphics[width=\linewidth]{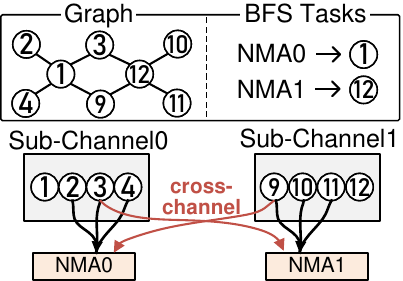}
    \vspace{-0.8em}
    \caption{Cross-channel access}
    \label{fig:cross_channel}
\end{subfigure}
    \caption{\textbf{(a) Latency breakdown} of ANNS-on-NDP design without \sysname optimizations; \textbf{(b) Cross-channel communication} highlighted in red when NMA0 and NMA1 perform BFS on node 1 and 12.}
\end{figure}

\subsection{Challenges of Deploying ANNS on NDP}
\label{sec:challenge_on_ndp_anns}

The execution of ANNS on NDP involves three steps:
\ballnumberA{1} Host CPU offloads distance calculation commands to NDP with locations of vector entries;
\ballnumberA{2} NMAs independently fetch vector data and compute distances;
\ballnumberA{3}
The host CPU gathers the results and looks up the neighbor lists to determine the next-hop vectors to visit.
\cref{fig:motivation_breakdown} shows the execution time breakdown of a vanilla ANNS.
The control overhead arises from \ballnumberA{1}, and the index lookup overhead from \ballnumberA{3}.
For \ballnumberA{2}, we further break the latency into distance computation and cross-channel memory access, and identify the following challenges:

\subsubsection{Overhead of distance calculations}
\label{subsec:moti_ndp_dist_cal}

\begin{figure}[t]
    \centering
    \includegraphics[width=\linewidth]{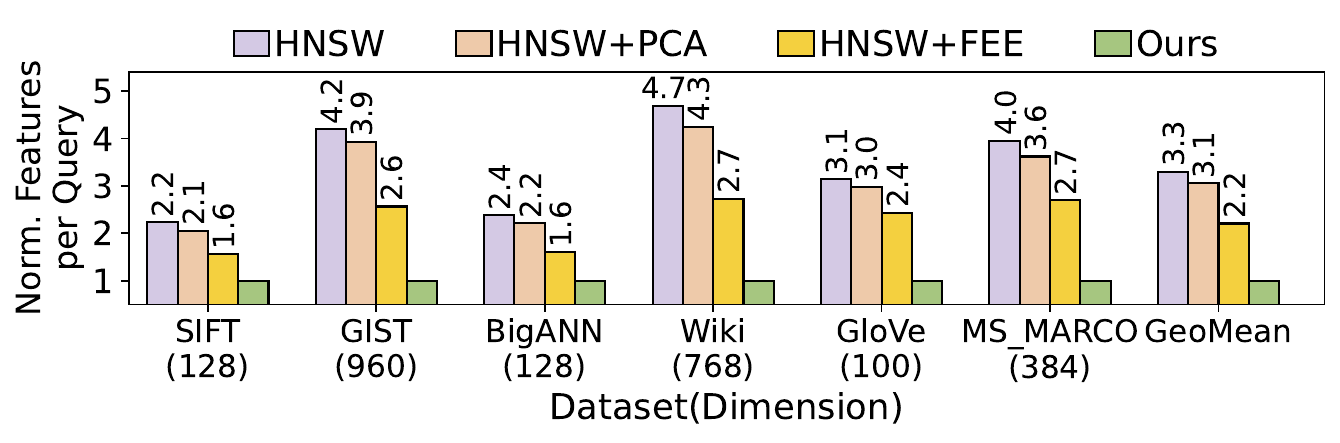}
    \caption{
    \textbf{Feature usage of HNSW variants on different datasets}, for algorithms achieving $\textrm{recall}@10 > 90\%$.
    }
    \label{fig:feature_usage_analysis}
\end{figure}

As shown in \cref{fig:motivation_breakdown}, distance computation dominates ANNS-on-NDP latency, particularly for GIST with 960-dimensional features.
\textit{This overhead can be reduced by lowering the number of features computed per vector}.
Prior optimizations mainly include principal component analysis (PCA) \cite{PCA2024ANNS} and feature-level early exiting (FEE) \cite{li2025ansmet,gao2023adsampling,gao2024rabitq}.
However, as shown in \cref{fig:feature_usage_analysis} under $\mathrm{recall}@10 > 90\%$, naive PCA reduces feature usage by only 6\%, and existing FEE methods (\cref{sec:fee_bg}) still leave considerable redundant computation.

\textbf{Our solution} approaches the problem from two aspects:
(1) Further reduce the number of features involved in early exiting; and
(2) Increase the number of features that can be fetched by each NDP data burst access.
For (1), we optimize FEE by comparing the $threshold$ with $d^k_\textrm{est}$ instead of $d^k_\textrm{part}$.
We propose FEE-sPCA in \cref{sec:FEE} to estimate $d_\textrm{all}$ based on $d^k_\textrm{part}$ while maintaining search accuracy.
Since $d^k_\textrm{est}\geq d^k_\textrm{part}$, $d^k_\textrm{est}$ between a query and a node can exceed the $threshold$ earlier, thereby triggering the FEE more promptly than using $d^k_\textrm{part}$.
For (2), we propose a dynamic floating-point (Dfloat) representation in \cref{sec:DFloat}, using variable bit-width for exponent and mantissa without hampering the search accuracy.
Thus, each DRAM burst can contain more features.

\subsubsection{Cross-channel memory accesses}
\label{subsubsec:moti_cross_channel}

\cref{fig:motivation_breakdown} shows that memory access overhead on NDP is also significant.
This overhead arises when an NMA must compute the distance of a vector stored in another sub-channel, incurring costly cross-channel access, as discussed in \cref{sec:ndp}.
The root cause is poor data locality in the graph structure.
As illustrated in \cref{fig:cross_channel}, when NMA0 performs the BFS of \ballnumberB{1}, it must access neighbors \ballnumberB{2}\ballnumberB{3}\ballnumberB{4}\ballnumberB{9}.
Since \ballnumberB{9} resides in a different sub-channel, the access incurs expensive cross-channel communication.
A similar issue occurs when NMA1 accesses the neighbors of \ballnumberB{12}.

\textbf{Our solution} proposes data-aware neighbor list mapping (DaM) in \cref{subsec:nbr_list_mapping}.
Following the vector data mapping across sub-channels, we also distribute the neighbor list to ensure that neighbor indices and vector data are resident in the same sub-channel, avoiding cross-channel data fetches.

\subsubsection{Costly CPU usage in naive ANNS-on-NDP}
\label{subsubsec:moti_CPU_esage}

\cref{fig:motivation_breakdown} also shows that CPU-side neighbor-list lookup contributes a significant fraction of the total latency.
This step lies on the critical path of ANNS-on-NDP, because NDP devices must wait for the CPU to identify the next-hop neighbors before launching the next round of distance computations.
Prior ANNS-on-NDP works largely overlook this overhead \cite{li2025ansmet,zeng2023dfgas,Bing2024SmartSSDs,Junhyeok2023CXL-ANNS,hu2023ice}, but our profiling shows that it accounts for about 31.7\% of total latency, mainly due to duplicated neighbor-list accesses.

\textbf{Our solution} also offloads neighbor-list lookup to NDP to exploit internal parallelism and bandwidth based on DaM.
We further incorporate a custom local neighbor cache (LNC) in \cref{subsec:local_neighbor_cache}, which stores recently accessed neighbor lists to avoid redundant accesses.

\section{Compressing Vector Database with VD-Zip}

\label{sec:VD-Zip}

For ANNS acceleration, we propose a software solution called \emph{VD-Zip} to compress the VecDB,
consisting of a feature-level optimization (FEE-sPCA) and a bit-level optimization (Dfloat).
During offline preprocessing, FEE-sPCA first applies a PCA transformation to the vector database,
enabling the effective estimation of full distance based on partial distance.
We further employ a statistical method to refine the estimation, ensuring a high recall rate.
During the online searching, the estimation is used to trigger FEE earlier.
Dfloat further lowers the DRAM data access by compressing more features within a single burst
while maintaining a high recall rate.

\begin{figure}[t]
    \centering
    \includegraphics[width=\linewidth]{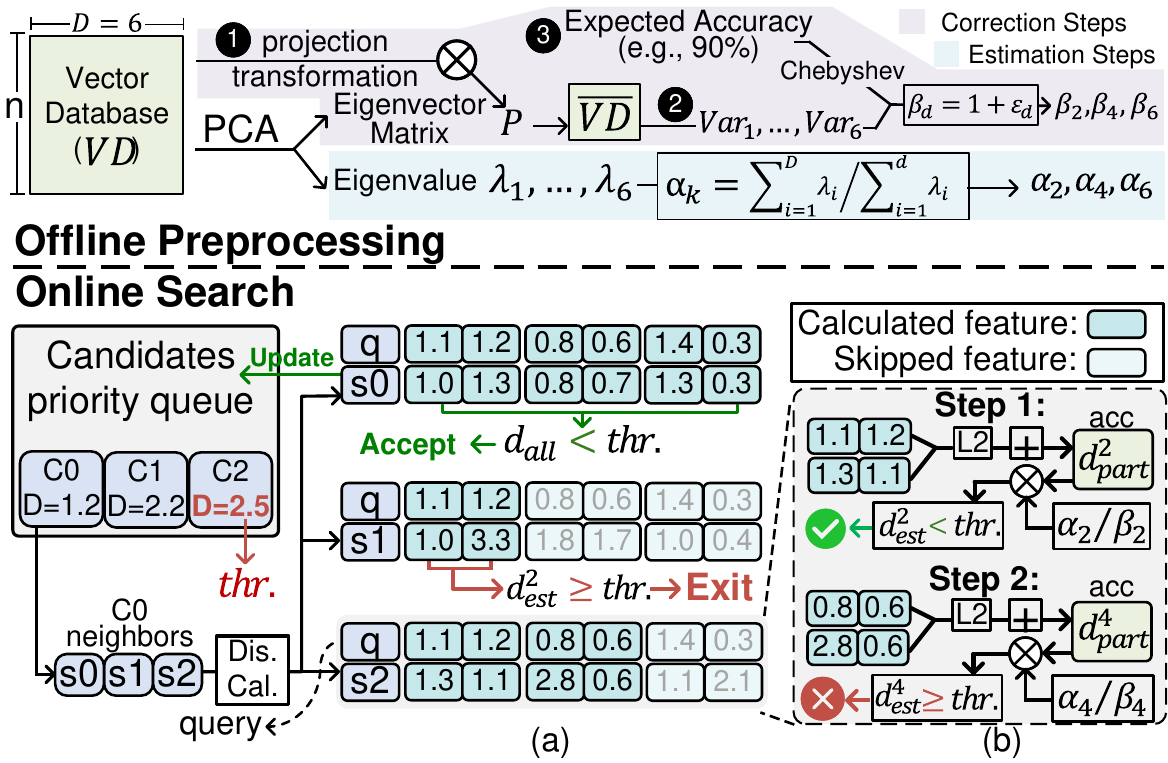}
    \caption{\textbf{FEE-sPCA execution flow}, including offline preprocessing (upper part) and online search
    (lower part).
    (a) Three neighbors (\ie, $\bm{s}0,\bm{s}1,\bm{s}2$) are searched and
    only $\bm{s}0$ is updated into priority queue. Computations of $\bm{s}1$ and $\bm{s}2$ are early exited.
    (b) Detailed steps of FEE-sPCA on $\bm{s}2$.}
    \label{fig:FEE}
    \vspace{-1.5em}
\end{figure}

\subsection {Feature-Level EE with Statistics-based PCA}
\label{sec:FEE}

The primary objectives of FEE-sPCA are
(1) leveraging estimated distance ($d^k_\textrm{est}$) to filter out non-candidate vectors (whose $d_\textrm{all}$ $\geq$ $threshold$) with partial distance ($d^k_\textrm{part}$); and
(2) controlling the accuracy of estimated distances to avoid erroneously filtering out candidate vectors (whose $d_\textrm{all}< threshold$).
To meet the two goals, we set two sets of parameters
$\alpha=\{\alpha_k\}$ and $\beta=\{\beta_k\}$ in the distance calculation process.
$\alpha_k$ is used to estimate the distance based on $d^k_\textrm{part}$ for (1).
$\beta_k$ is used to calibrate the estimation to maintain accuracy for (2).
The overall process, including the offline pre-processing
(to obtain $\alpha$, $\beta$) and online searching, is shown in \cref{fig:FEE}.

In this subsection, we first introduce the online searching
flow with our FEE-sPCA (lower part in \cref{fig:FEE}).
Then we introduce how the parameters $\alpha_k$ and $\beta_k$ are determined offline (upper part in \cref{fig:FEE})
for computing $d^k_{\textrm{est}}$.

\subsubsection{Online searching with FEE-sPCA}
The lower part of \cref{fig:FEE} shows the process.
The candidate priority queue stores the identified nearest candidates $\{\vc0, \vc1, \vc2\}$ of current query $\vq$, and their distances \wrt $\vq$ are \{1.2, 2.2, 2.5\}.
$\{\bm{s}0, \bm{s}1, \bm{s}2 \}$ are neighbors of the queue head $\vc0$, and they are the new nodes to be searched in this hop.
The process is to calculate the distance between $\{\bm{s}0, \bm{s}1, \bm{s}2 \}$ and $\vq$, then update the vector in the candidate priority queue if its distance $\leq$ threshold (2.5).
As shown in \cref{fig:FEE}a, we assume each DRAM access can get 2 features, so we calculate the distance of 2 dimensions each time.
Only $\bm{s}0$ is accepted and updated in the queue, while the calculations of $\bm{s}1$ and $\bm{s}2$ are terminated with FEE.
The calculation of $\bm{s}1$ exits after the first 2 features are calculated, while $\bm{s}2$ exits after the first 4 features.

Taking $\bm{s}2$ as an example to describe the FEE-sPCA, as shown in \cref{fig:FEE}b, each DRAM burst corresponds to one step (\eg, Step 1/2), loading 2 dimensions.
In Step 1, it calculates the partial distance of the first two features ($d^2_\textrm{part}$) between $\vq$ and $\bm{s}2$.
Then, we obtain the estimated distance $d^2_\textrm{est}=\alpha_2\cdot d^2_\textrm{part}/\beta_2$.
We compare the estimated distance $d^2_\textrm{est}$ with $threshold$.
As $d^2_\textrm{est} < threshold$, we proceed to Step 2 to calculate the next two features' distance and accumulate it to the last calculated $d^2_\textrm{part}$ to get a new partial distance $d^4_\textrm{part}$.
Based on $d^4_\textrm{part}$, we update the estimated distance $d^4_\textrm{est}$.
As $d^4_\textrm{est} \geq threshold$, early exiting is triggered.

\subsubsection{Offline preprocessing via PCA to get $\alpha$}
\label{subsec:offline_pca_alpha}
We aim to get $d^k_\textrm{est}$ based on the partially computed $d^k_\textrm{part}$ from the first $k$ dimensions.
To address this, we preprocess the database offline as shown in \cref{fig:FEE} upper part (blue).
We first apply PCA to make the leading dimensions of all vectors contain the most informative components.
As PCA is a linear dimensionality reduction technique, it can be effectively applied to these vectors, which are approximately linear after the embedding transformation \cite{arora2018linear,Mikolov2013nips,Mikolov2013ohu}.
After PCA, in addition to the generation of eigenvalue $\lambda_i$ for each dimension and one eigenvector matrix $\mP$, there exists an expectation property of:
\begin{equation}
E\left (  \left \| \vv_{1:d} \right \|^2 /\left \| \vv \right \|^2   \right ) =
{\textstyle \sum_{i=1}^{d}\lambda_i}/{\textstyle \sum_{i=1}^{D}\lambda_i}
\label{eqt:pca_expection}
\end{equation}
where $\vv$ is a vector in the transformed VecDB $\overline{VD}$,
and $\left \| \vv \right \|^2$ is the squared norm of all its features.
$\vv_{1:d}$ contains the first $d$ features.
$\lambda_i$($1 \le i\le D$) is the eigenvalue of the $i$-th feature, obtained by the PCA process offline.
Then we can get:
\begin{equation}
 d_\textrm{all}  \approx
d^k_\textrm{part} \cdot
{\textstyle \sum_{i=1}^{D}\lambda_i}/{\textstyle \sum_{i=1}^{k}\lambda_i}
\label{eqt:pca_estimate}
\end{equation}
We make the parameter $\alpha_k = {\textstyle
\sum_{i=1}^{D}\lambda_i}/{\textstyle \sum_{i=1}^{k}\lambda_i}$.
Therefore, $d_\textrm{all} \approx d^k_\textrm{est}=\alpha_k \cdot d^k_\textrm{part}$.
However, the estimation \textbf{may cause errors} in FEE.
Thus, we further propose a correction.

\begin{figure}[t]
    \centering
    \includegraphics[width=\linewidth]{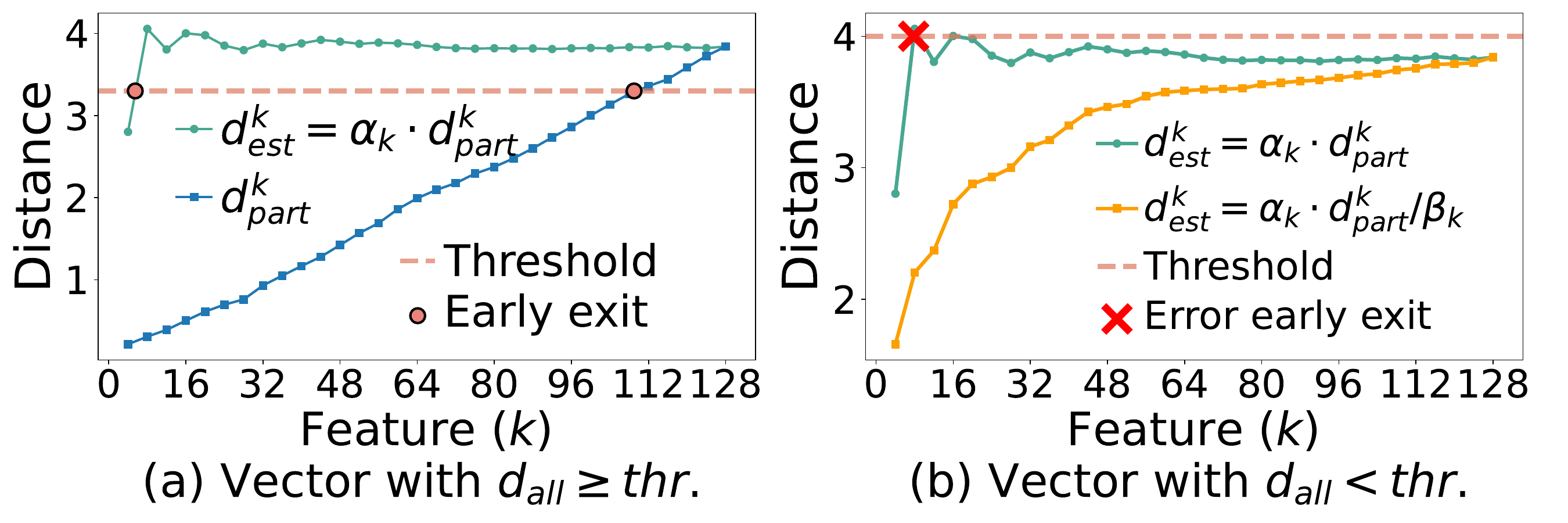}
    \vspace{-1.5em}
    \caption{\textbf{Calculated distance versus used features} and its relationship to the threshold. Data is from SIFT1M.}
    \vspace{-1em}
    \label{fig:FEE_reject_accept}
\end{figure}

\subsubsection{Offline preprocessing to get $\beta$}

In \cref{fig:FEE_reject_accept}, we present two examples that illustrate the need for correction.
In \cref{fig:FEE_reject_accept}a, the vector satisfies $d_{\textrm{all}} \geq \mathit{threshold}$ and should be rejected.
Its partial distance $d_{\textrm{part}}^{k}$ triggers FEE at the 109th feature, whereas the estimated distance $d_{\textrm{est}}^{k}$ triggers FEE much earlier at the 4th feature, showing higher FEE effectiveness.
However, \cref{fig:FEE_reject_accept}b shows a vector with $d_{\textrm{all}} < \mathit{threshold}$ that should be accepted.
Using only the PCA-based estimate, $d_{\textrm{est}}^{k}=\alpha_k \cdot d_{\textrm{part}}^{k}$, incorrectly triggers FEE at around the 8th dimension because $d_{\textrm{est}}^{8}$ overestimates the distance and exceeds $\mathit{threshold}$.
To preserve search accuracy, such false rejections must be minimized.
We therefore scale down $d_{\textrm{est}}^{k}$ by dividing it by a factor $\beta > 1$, reducing overestimation and preventing erroneous early exits.
The corrected estimate is the yellow dotted line in \cref{fig:FEE_reject_accept}b.

The following description shows the procedure to acquire $\beta$.
We first analyze the property of the estimation error between $d^k_\textrm{est}$ and $d_\textrm{all}$.
Based on \cref{eqt:pca_expection}, we can get:
\begin{equation}
E\left ( \alpha_k \cdot d^k_\textrm{part} / d_\textrm{all}   \right ) = 1
\label{eqt:exception}
\end{equation}
Furthermore, each $d^k_\textrm{part}$ has its own variance.
Therefore, we can apply Chebyshev’s inequality to $\alpha_k \cdot d^k_\textrm{part} / d_\textrm{all}$:
\begin{equation}
P( \left | \alpha_k \cdot d^k_\textrm{part} / d_\textrm{all} -1 \right | \le \varepsilon _k) \ge 1-{Var_k}/{\varepsilon _k^2}
\label{eqt:pca_exp_cheby}
\end{equation}
where $P$ is the probability, $\varepsilon_k$ is a tiny positive number, and $Var_k$ is the variance of $
\alpha_k \cdot d^k_\textrm{part} / d_\textrm{all}$, which can be obtained during index construction.
After removing the absolute value and letting $1+\varepsilon_k=\beta_k$:
\begin{equation}
P\left ( \alpha_k \cdot d^k_\textrm{part}/ \beta_k  <  d_\textrm{all}  \right ) \ge 1- {Var_k}/{2\varepsilon_k ^2}
\label{eqt:pca_exp_cheby_final}
\end{equation}
To ensure that $d^k_\textrm{est}\leq d_\textrm{all}$ with high probability to avoid FEE errors,
we can make $1- Var_k/2\varepsilon_k ^2$ a
large value (\eg, 90\%) and get the corresponding $\varepsilon_k$ and $\beta_k$.
The flow is shown in \cref{fig:FEE} upper purple part, which begins by projecting the database to obtain $\overline{VD}$ and its variance (\ballnumberA{2}).
We set an expected accuracy (\ballnumberA{3}), and obtain $\beta_k$ by using \cref{eqt:pca_exp_cheby_final}.
As shown in \cref{fig:FEE_reject_accept}b, after the adoption of the statistical method with $\beta_k$, the sPCA $d^k_\textrm{est}$ is corrected and avoids the FEE error.

\subsubsection{Result}
\begin{figure}[t]
    \centering
    \includegraphics[width=\linewidth]{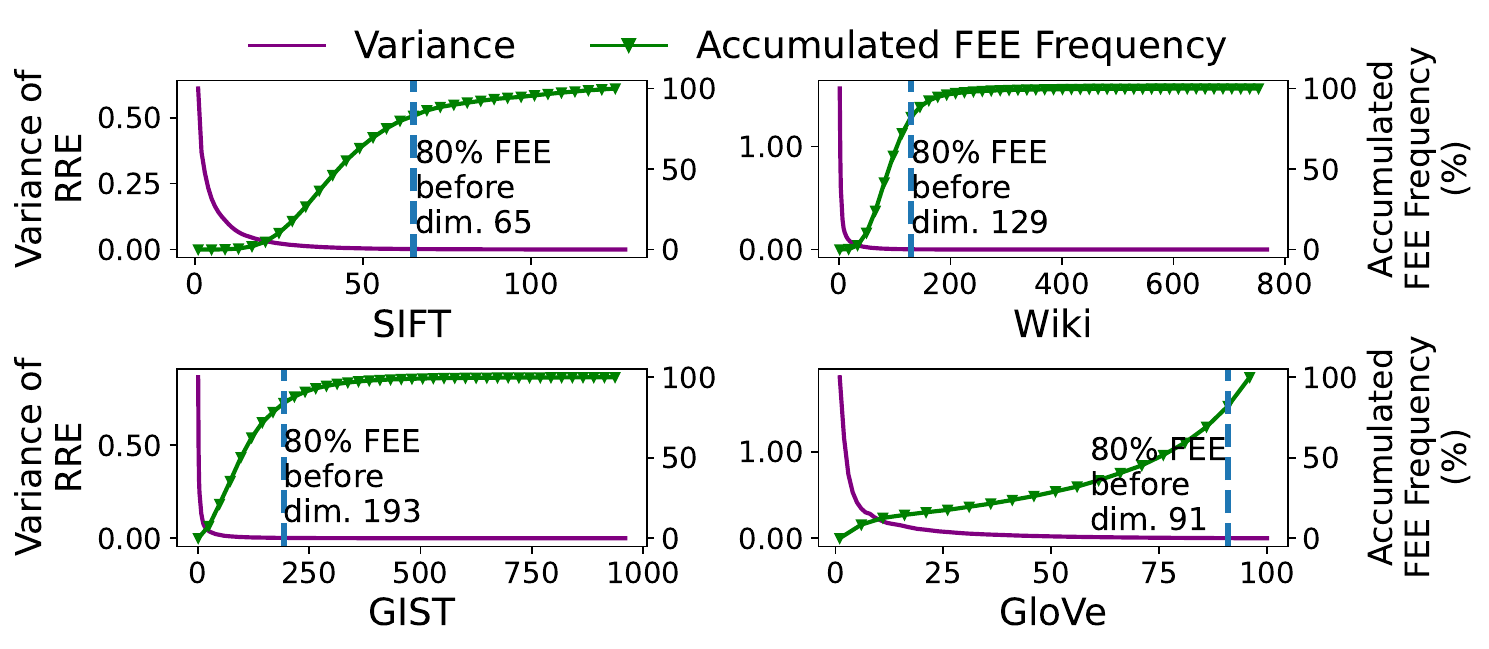}
    \vspace{-1.5em}
    \caption{
    \textbf{Performance of FEE-sPCA}. The purple line denotes the variance term in \cref{eqt:pca_exp_cheby_final}, the green line denotes the dimension-wise accumulated FEE-sPCA trigger frequency, and the dashed line denotes the dimension before which 80\% of computations terminate.
    }
    \vspace{-1em}
    \label{fig:V_vs_Freq}
\end{figure}

We further present the $Var_k$ in \cref{eqt:pca_exp_cheby_final} and
results of the FEE-sPCA technique across datasets in \cref{fig:V_vs_Freq}, covering the dimension from 128 to 960 and
including L2 and IP distance.
Overall, we can evenly reduce feature calculations by nearly 50\%,
especially for high-dimensional datasets
(\eg 80\% FEEs are triggered within the first 193 dimensions on the GIST dataset with 960 dimensions
per vector).

\subsection{NDP-Aware Dynamic Floating-Point Representation}
\label{sec:DFloat}

We introduce dynamic floating-point (Dfloat) to reduce the number of bits per feature, thereby increasing the number of features retrieved per DRAM burst.
Conventional low-precision formats (e.g., BF16/FP16/FP8) are not well suited for FEE-sPCA because they quantize all dimensions uniformly.
After applying FEE-sPCA, however, different dimensions contribute unequally, and uniform quantization noticeably degrades its robustness and accuracy.
We therefore propose Dfloat, which provides a more robust representation tailored to the characteristics of FEE-sPCA.

\subsubsection{Representation of Dfloat}
Lowering the bit width of vector features is an effective approach to reduce the memory footprint and data movement.
In this work, we leverage the dynamic floating-point representation
(Dfloat) \cite{tambe2020algorithm,liu2021improving} with adaptive bit widths
for the exponent and mantissa,
\ie,
\begin{multline}
\label{eqt:dfloat-normalized}
g(\vb_\textrm{dfloat}) =
\eqnmarkbox[red]{node1}{ (-1)^{b_{ n_\textrm{exp}+n_\textrm{man} }} } \times
\eqnmarkbox[gray]{node2}{
2^{\sum_{i= n_\textrm{man}}^{ n_\textrm{exp}+n_\textrm{man}-1 } 2^{i- n_\textrm{man}}\cdot b_i - B}
}
\\
\times
\eqnmarkbox[blue]{node3}{
\left(1+ \displaystyle \sum_{i=0}^{n_\textrm{man}} 2^{(i-n_\textrm{man})} \cdot b_i\right)
}
;
\quad
\vb_\textrm{dfloat} = \{b_i\}_{i=0}^{n_\textrm{exp}+n_\textrm{man}}
\annotate[xshift=-4em,yshift=.5em]{below,right}{node1}{sign}
\annotate[xshift=2em, yshift=.6em]{below,right,label above}{node2}{exponent}
\annotate[xshift=-7em, yshift=.6em]{below}{node3}{mantissa}
\end{multline}
where $\vb_\textrm{dfloat}$ is the binary representation with $b_i \in \{0,1\}$.
$n_\textrm{exp}$ and $n_\textrm{man}$ are the bit widths of the exponent and mantissa.
We introduce NDP-aware optimization to Dfloat for our system.

\subsubsection{NDP-aware optimization}
\begin{figure}[t]
    \centering
    \includegraphics[width=.9\linewidth]{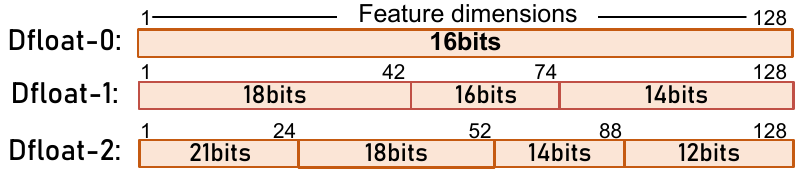}
    \vspace{-.5em}
    \caption{\textbf{Example Dfloat configurations.}
        Features are divided into segments with different
    $\textrm{bit width}=1+n_\textrm{exp}+n_\textrm{man}$.}
    \label{fig:sample_dfloat_config}
\end{figure}

Based on our preliminary results, simply applying one configuration
(\ie, small $n_\textrm{exp}$ and $n_\textrm{man}$) for all dimensions
leads to a significant recall degradation.
It occurs mainly because our sPCA transformation concentrates more important information in lower dimensions, and those dimensions are more sensitive to the low bit-width representation.
To achieve better bit-level compression, \emph{we propose to conduct a fine-grained search to identify an optimized Dfloat configuration,
to maximize ANNS throughput and recall rate.}
We first divide a vector into $N_\textrm{seg}$ segments along the feature
dimension,
each with a different bit width.
We formulate the optimization objective to minimize the number of DRAM bursts for accessing one vector $N_\textrm{burst}$ while keeping the ANNS recall rate above a preset
threshold $R_\textrm{target}$:
\begin{equation}
    \min_{ \sC_\textrm{opt} } N_\textrm{burst} ; \quad \textrm{Subject to:}~ R(\sC_\textrm{opt}) > R_\textrm{target}
\end{equation}
where $R(\sC_\textrm{opt})$ is the recall evaluated when VecDB is processed
with an optimized Dfloat configuration $\sC_\textrm{opt} = \{ n_{\textrm{exp},i},
n_{\textrm{man},i}  \}_{i=1}^{N_\textrm{seg}}$.
Taking configuration \textbf{Dfloat-1} in \cref{fig:sample_dfloat_config} as an
example, the vector is divided into three segments via a search algorithm ($N_\textrm{seg} = 3$).
The chosen Dfloat configuration for the 1st segment is
$1 + n_{\textrm{exp},1}+ n_{\textrm{man},1} = 18$.

\begin{algorithm}[t]
\caption{Search algorithm for Dfloat configuration.}
\label{alg:dfloat-dse}
\begin{algorithmic}[1]
    \State \textbf{Input:} Target recall@k = $R_\textrm{target}$;
    Number of features each vector = $d$;
    Recall@k with subsets of queries = $R'(\cdot)$, $1+n_\textrm{exp} + n_\textrm{man} \in [12,32]$;
    Number of bits per burst $B_\textrm{burst}$
    \State \textbf{Output:} Optimized Dfloat configuration $\sC_\textrm{opt}$
\end{algorithmic}
\begin{algorithmic}[1]
    \State $N_\textrm{burst}^\textrm{max}  \gets d/(B_\textrm{burst}/32); \quad N_\textrm{burst}^\textrm{min}  \gets d/(B_\textrm{burst}/12)$
    \While{$N_\textrm{burst}^\textrm{min} < N_\textrm{burst}^\textrm{max}$}
        \State $N_\textrm{burst} = \floor{ (N_\textrm{burst}^\textrm{min} + N_\textrm{burst}^\textrm{max}) /2}$ \textcolor{blue}{\Comment{Number of bursts}}
        \State $ \{\sC\} \gets \textrm{cfg-validate} ( N_\textrm{burst}) $ \textcolor{blue}{\Comment{All valid configs}}
        \For{$i=1$ to \textrm{\#configs}($\{\sC\}$) $\& ~\sC \neq \emptyset$  }
            \If{ $R(\sC_i) \ge R_\textrm{target}~\& ~R(\sC_i) > R(\sC_\textrm{opt})    $ }
                \State $N_\textrm{burst}^\textrm{min} \gets N_\textrm{burst}$; $\sC_\textrm{opt} \gets \sC_i$;
            \EndIf \EndFor
        \If{$N_\textrm{burst}^\textrm{min} \neq N_\textrm{burst}$}
            \State $ N_\textrm{burst}^\textrm{max} \gets N_\textrm{burst}$
        \EndIf
    \EndWhile
    \State \textbf{Return} $\sC_\textrm{opt}$; \textcolor{blue}
    {\Comment{$\{n_\textrm{exp}, n_\textrm{man}\}$ for each vector segment}}
\end{algorithmic}
\end{algorithm}

\begin{figure*}[t]
    \centering
    \includegraphics[width=\textwidth]{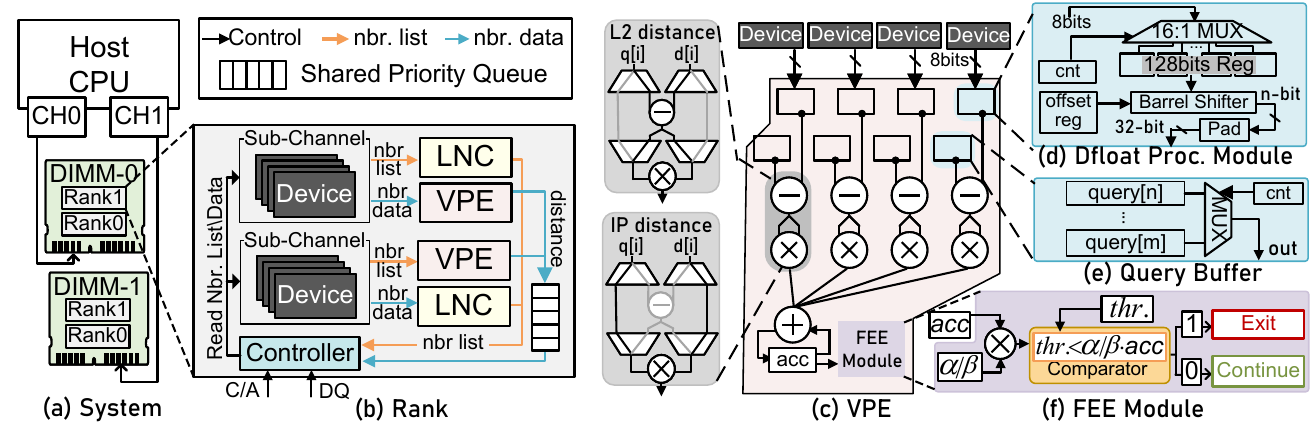}
    \vspace{-1.5em}
    \caption{\textbf{Hardware architecture overview of \sysname.} The host CPU connects to DIMM-based DRAM modules via memory channels, where each rank embeds near-memory hardware.
}
    \label{fig:overall}
    \vspace{-1em}
\end{figure*}

Given a specific VecDB,
to identify an optimized Dfloat configuration $\sC_\textrm{opt}$,
we combine binary search and brute force enumeration as described in \cref{alg:dfloat-dse}.
The general idea of \cref{alg:dfloat-dse} is performing the binary search between max- and min-bound of possible $N_\textrm{burst} \in [N_\textrm{burst}^\textrm{min}, N_\textrm{burst}^\textrm{max}]$.
For a specific $N_\textrm{burst}$, we conduct an exhaustive search and filter out all possible
Dfloat configurations via validation (line-4 in \cref{alg:dfloat-dse}) following
the rules:
\begin{enumerate}[leftmargin=*]
    \item Features of one DRAM burst use identical Dfloat format;
    \item When the number of features per burst is set, we are prone to increase Dfloat bit width to achieve higher recall;
    \item The feature bit width ($1+n_\textrm{exp} + n_\textrm{man}$) gradually decreases with the feature index increasing;
    \item $N_\textrm{burst}$ must be a multiple of the number of devices per sub-channel,
        as devices work synchronously.
\end{enumerate}
Note that the DRAM burst size ($B_{\textrm{burst}}$) depends on the DDR generation, \eg, 128 bits for DDR5 and 64 bits for DDR4.

Line 6 of \cref{alg:dfloat-dse} evaluates several sampled queries to characterize the database through multiple searches.
To ensure broad coverage of HNSW traversal paths and avoid repeatedly probing localized regions, the sampled queries should be diverse. We select them from the full train set of benchmark or sample 1K queries from test set if train set is absent, which is sufficiently representative and covers most index paths.
To efficiently explore the Dfloat design space, we use a mask-based emulation method on the host CPU: by applying bit masks to 32-bit floating-point data, we emulate the precision loss of different configurations without repeatedly rebuilding the index.
For frequently updated databases, we run the offline process (including both FEE-sPCA and Dfloat) only when updates reach about 30\% of the database, at which point the vector index itself typically also requires rebuilding due to structural degradation.

\subsubsection{Portability}
Dfloat improves performance only by increasing the number of features retrieved per memory access, without changing the computation itself or requiring specialized computation units.
Before entering the FPU, Dfloat values are zero-padded to match standard arithmetic units (FP32 in \sysname).
Dfloat packing is performed offline during pre-processing.
It is independent of any particular floating-point format and can be applied to existing floating-point representations.

\subsubsection{ECC Compatibility}
Server-grade DDR5 DIMMs typically use both on-die ECC and side-band ECC for reliability~\cite{on-die-ecc,jedec2020ddr5}.
In on-die ECC, DRAM chips internally compute the ECC for the written data and store the ECC code.
Since \sysname adds NMA logic in a separate chip without modifying DDR5 dies, on-die ECC remains unaffected.
As for side-band ECC, it has additional DRAM chips for ECC bits storage.
However, Dfloat is only a software-level data representation, and the physical DRAM chips still follow the standard DDR5 burst format.
Thus, conventional memory-controller ECC correction~\cite{ddr-ecc} remains compatible with \sysname.

\section{Hardware Architecture}

\label{sec:arch}

\subsection{Architecture Overview}
\label{sec:overall_arch}

The overall architecture of \sysname is shown in \cref{fig:overall}, consisting of a host CPU and multiple DIMM-based DDR5 DRAM modules connected via memory channels.
An example configuration is illustrated in \cref{fig:overall}a, where two memory channels are each connected to one DIMM.
Each DIMM contains multiple ranks and incorporates customized hardware to accelerate ANNS.

\cref{fig:overall}b shows the micro-architecture of a rank, in which DRAM devices are organized into two DDR5 sub-channels.
Each sub-channel contains four DRAM devices with 8-bit IO width.
\sysname integrates a vector processing engine (VPE) and a local neighbor cache (LNC) into each sub-channel for efficient near-memory ANNS acceleration.
The VPE computes distances for vectors retrieved from the local sub-channel, while the LNC caches frequently accessed neighbor lists to reduce redundant memory accesses.
A shared priority queue after the two VPEs merges and sorts their results, so that only top candidates are returned to the host CPU, reducing both data transfer and CPU-side overhead.
The controller, shared priority queue, two LNCs, and VPEs are packaged together with the buffer chip.
We next describe the VPE design in \cref{subsec:vector_process_engine}, followed by our data-aware mapping (DaM) and local neighbor cache (LNC) designs in \cref{subsec:data_and_nbr_list_mapping} and \cref{subsec:local_neighbor_cache}, respectively.

\subsection{Vector Process Engine}
\label{subsec:vector_process_engine}

\cref{fig:overall}c shows the microarchitecture of the VPE, which integrates the FEE and Dfloat optimizations described in \cref{sec:FEE} and \cref{sec:DFloat}.
The VPE contains four parallel processing paths, each corresponding to one DRAM device.
Each path includes a Dfloat processing module, a query buffer, and a distance calculation module.
The outputs of the four paths are then merged by an accumulator, whose result dynamically guides the FEE module to trigger early exit.

The \textit{Dfloat process module}, shown in \cref{fig:overall}d, decodes Dfloat-formatted vector data retrieved from the DRAM device.
DRAM data are read in bursts, with each device supplying 128 bits per burst, \ie, 8 bits per cycle over 16 cycles.
Accordingly, a counter-controlled 16-to-1 multiplexer sequentially loads the 16 bytes of a burst from one DRAM chip into a 128-bit register.
Once the register is filled, a barrel shifter extracts each \(n\)-bit Dfloat element according to the preset offset register.
The extracted value is then zero-padded to 32-bit floating point, completing the decoding process.

The \textit{query buffer}, shown in \cref{fig:overall}e, stores query vector elements preloaded by the CPU before search.
During computation, a wrapped-counter-driven multiplexer sequentially outputs one query element per cycle for distance calculation.

The \textit{distance calculation module} (gray-highlighted in \cref{fig:overall}c) supports both L2 distance and inner-product (IP) computation between the query and vector data.
It adopts a shared datapath with a multiplexer to switch between the two modes, following prior designs \cite{li2025ansmet, yuan2025fanns, Junhyeok2023CXL-ANNS}.
The partial distances produced by the four parallel modules are then accumulated in the accumulator.

The \textit{FEE module}, shown in \cref{fig:overall}f, determines whether early exit should be triggered.
Whenever the accumulator is updated, the module estimates the final distance by scaling the current partial sum with factors $\alpha_k$ and $\beta_k$, following \cref{sec:FEE}.
The estimation is then compared with the threshold, \ie, the distance of the current farthest point in the candidate queue.
If the estimated distance exceeds the threshold, early exit is triggered and the vector is discarded.

\subsection{Mapping of Data and Neighbor List}
\label{subsec:data_and_nbr_list_mapping}

\begin{figure}[t]
    \centering
    \includegraphics[width=.9\linewidth]{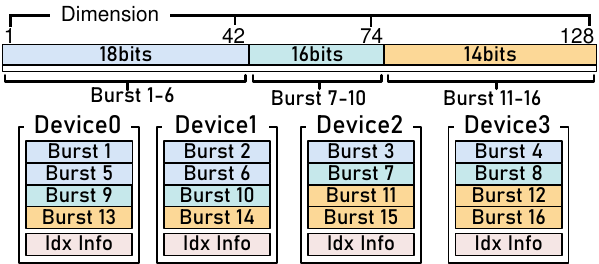}
    \vspace{-.5em}
    \caption{\textbf{An example 128-dimensional vector data mapping within a sub-channel} (on SIFT \cite{lowe2004sift} dataset).}
    \label{fig:nbr_vector_data}
    \vspace{-1em}
\end{figure}

\subsubsection{Data mapping}
\label{subsec:mapping}

\sysname maps each vector entirely to a single sub-channel, with its dimensions distributed across the four DRAM devices.
\cref{fig:nbr_vector_data} shows an example for a 128-dimensional vector.
With Dfloat encoding, dimensions 1$\sim$42, 43$\sim$74, and 75$\sim$128 are assigned 18, 14, and 16 bits, respectively.
Since each device provides 128 bits per burst, these three segments require six, four, and six bursts, respectively.
The bursts are interleaved across the four devices, so that in each memory access all devices return one burst in parallel.
Access then proceeds sequentially until all dimensions are processed, naturally matching FEE, which evaluates dimensions in increasing order.

\begin{figure}[t]
    \centering
    \includegraphics[width=.85\linewidth]{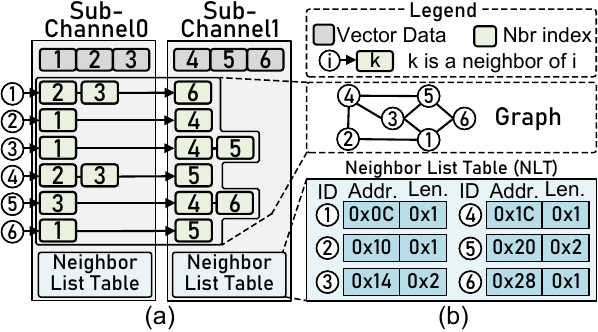}
    \vspace{-.75em}
    \caption{
    \textbf{Data-aware neighbor list mapping} (DaM).
    Neighbor lists are partitioned across sub-channels.}
    \label{fig:nbr_list_data}
    \vspace{-1em}
\end{figure}

\subsubsection{Data-aware neighbor list mapping (DaM)}
\label{subsec:nbr_list_mapping}

\sysname stores neighbor lists on NDP to offload neighbor retrieval from the CPU to NDP, as discussed in \cref{subsubsec:moti_cross_channel}.
To reduce cross-channel communication and enable parallel lookup, \sysname places each neighbor list in a data-aware manner, co-locating it with the corresponding vector in the same sub-channel.
As a result, each sub-channel can independently retrieve neighbors and compute distances for its local vectors, minimizing cross-channel data movement.

\cref{fig:nbr_list_data} shows an example with six nodes, where both vector data and partitioned neighbor lists are distributed across sub-channels.
For example, vector 1 has neighbors $\{2,3,6\}$.
Because vectors 2 and 3 are stored in sub-channel 0 while vector 6 is stored in sub-channel 1, the neighbor list of vector 1 is partitioned accordingly across the two sub-channels.
When the CPU issues a request to traverse the neighbors of vector 1, sub-channel 0 retrieves its local neighbor list and computes the distances for vectors 2 and 3.
In parallel, sub-channel 1 independently handles vector 6, eliminating the need for inter-sub-channel communication.

However, the length of each partitioned neighbor list differs across nodes, making efficient indexing nontrivial.
To address this, we store a neighbor list table (NLT) in each channel memory, as shown in \cref{fig:nbr_list_data}b.
The NLT records the length and memory address of the neighbor list, enabling efficient indexing of variable-length entries.
To further accelerate neighbor-list lookup, we also employ a local neighbor cache (LNC).

\begin{figure}[t]
    \centering
\includegraphics[width=\linewidth]{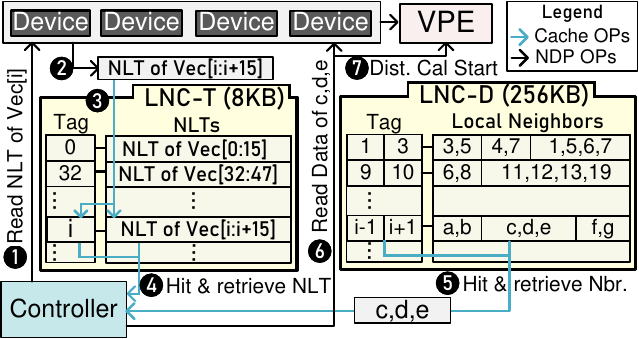}
    \vspace{-1em}
    \caption{\textbf{Illustration of local neighbor cache} (LNC). LNC-T caches entries of the Neighbor List Table (NLT), while LNC-D caches the actual neighbor list contents.}
    \label{fig:local_nbr_cache}
    \vspace{-1em}
\end{figure}

\subsection{Local Neighbor Cache}
\label{subsec:local_neighbor_cache}

The key insight is that neighbor-list accesses exhibit strong temporal and spatial locality: similar or repeated queries often revisit the same nodes, causing redundant lookups.
To exploit this locality, \sysname introduces the local neighbor cache for tables (LNC-T) and the local neighbor cache for data (LNC-D).
LNC-T stores NLT entries and functions similarly to a translation
lookaside buffer (TLB), while LNC-D stores the corresponding neighbor-list contents and functions like a data cache.
They together reduce memory accesses and improve search throughput.

\cref{fig:local_nbr_cache} illustrates the structure and operation of the LNC.
Its configuration is: LNC-T is an 8KB fully associative cache, while LNC-D is a 256KB 8-way set-associative cache.
Both use 64-byte cache lines, matching the burst size of a sub-channel.
The two caches use different tag formats.
Since each NLT entry (\cref{fig:nbr_list_data}b) occupies 4 bytes (3 bytes for the start address and 1 byte for the length), one LNC-T cache line stores 16 entries, so its tag only records the ID of the first entry.
By contrast, because neighbor-list sizes vary across sub-channels, the LNC-D tag records both the start and end node IDs of the cached neighbor-list segment.

\cref{fig:local_nbr_cache} also illustrates the LNC workflow.
Consider the distance calculation for vector $i$, where its NLT entry misses in LNC-T but its neighbor list hits in LNC-D.
The controller first requests the NLT entry of vector $i$ (\ballnumberA{1}), which is fetched from memory (\ballnumberA{2}) and inserted into LNC-T (\ballnumberA{3}).
The controller then reads the cached NLT entry from LNC-T (\ballnumberA{4}) to obtain the address of vector $i$'s neighbor list.
Next, the neighbor list is accessed from LNC-D with a cache hit (\ballnumberA{5}).
Using the length information (3 in this example), the controller identifies the local neighboring nodes as $c$, $d$, and $e$, and then issues requests to fetch their data and compute distances (\ballnumberA{6}, \ballnumberA{7}).

\begin{figure}[t]
    \centering
    \includegraphics[width=\linewidth]{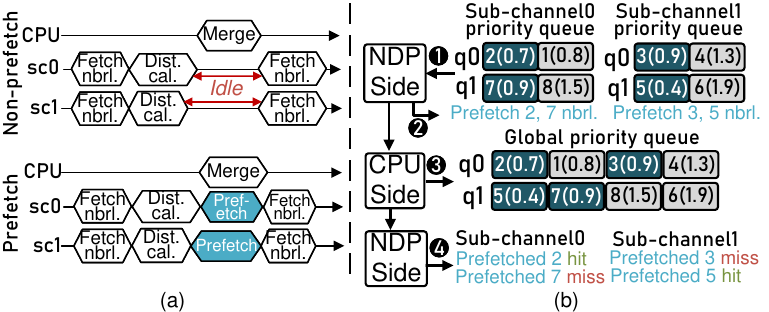}
    \caption{(a) \textbf{Comparison of flows} with and without prefetch. (b) \textbf{Execution flow with prefetch} under batch=2.}
    \vspace{-.5em}
    \label{fig:prefetch}
\end{figure}

\sethlcolor{hl_pink}
\begin{figure*}[t]
    \centering
    \includegraphics[width=.95\linewidth]{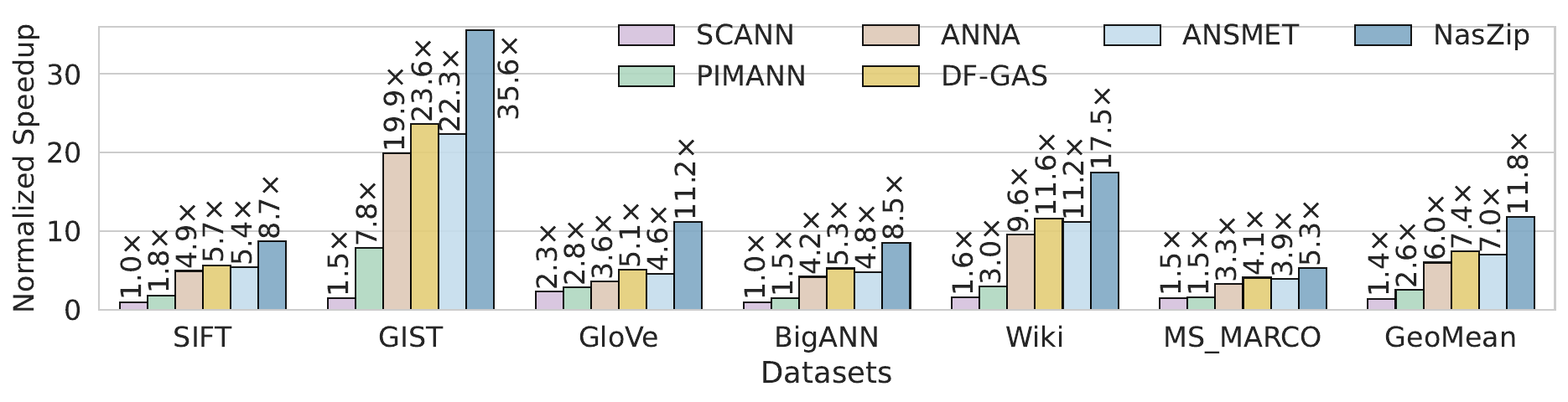}
    \vspace{-1em}
    \caption{\textbf{Throughput (QPS) across datasets} with recall@10 $\geq90\%$ on various architectures including CPU (SOTA SCANN), ASIC (ANNA), UPMEM (PIMANN), FPGA (DF-GAS), NDP (SOTA ANSMET, \sysname) normalized to CPU baseline.}
    \label{fig:eva_overall_QPS}
\end{figure*}
\sethlcolor{yellow}

\subsection{Neighbor List Prefetching and batch scheduling}
\label{subsec:prefetch}

Scheduling is performed by synchronizing the hop-by-hop traversal of multiple queries within a batch. Within each hop, vector distances are computed sequentially because our lightweight design provides only a limited number of FPUs, leaving little room for intra-hop scheduling optimization.
However, as shown by the baseline schedule in \cref{fig:prefetch}a upper part, we observe idle time between hops while waiting for CPU-side merging.
To exploit this gap, \sysname prefetches neighbor lists for the next BFS hop.
An example with two sub-channels and a batch of two queries is shown in \cref{fig:prefetch}b.
After each search hop (\ballnumberA{1}), the sub-channel priority queue stores the current vector IDs of each query (q0 and q1), ordered by distance. For example, 2(0.7) denotes vector 2 with distance 0.7.
Each sub-channel then prefetches the neighbor list (nbrl.) of the current closest vector for each query (\ballnumberA{2}) and stores the fetched data in the LNC.
Meanwhile, the queue contents are sent to the host CPU, which merges and sorts them in the global priority queue (\ballnumberA{3}).
The CPU then returns the global closest nodes to each sub-channel for the next BFS hop.
In the example, the next hop requires the neighbor lists of vectors 2 and 5, so sub-channel 0 and sub-channel 1 successfully reuse the prefetched data, respectively (\ballnumberA{4}).
As shown in \cref{fig:prefetch}a, this prefetching scheme fills the idle gap during CPU-side merging compared with the no-prefetch baseline.

When prefetching fails, the overhead remains small because the prefetched content is retained in the LNC (\cref{subsec:local_neighbor_cache}), where it can still be effectively reused by future accesses.
We further analyze the prefetch hit rate in \cref{subsubsec:eva_prefetch}, and the impact of batch size on scheduling in \cref{subsec:batch_eva}.

\section{Evaluation}

\subsection{Evaluation Methodology}
\label{sec:evaluation_methodology}

\subsubsection{Experimental setup}
We develop RTL implementations of added logic in \sysname.
The functionality of modules in \sysname is verified on FPGA.
For accurate area and power evaluation, the RTL design is synthesized using Synopsys Design Compiler with 28nm technology, and place-and-routed using Cadence Innovus.
System performance (QPS, latency, recall) is evaluated on UniNDP~\cite{xie2025unindp}, a cycle-accurate NDP simulator.
System configurations are specified in \cref{tab:sys_cfg}.

\begin{scriptsize}
\begin{table}[t]
\caption{\textbf{Evaluation Platforms and Configurations.}}
\label{tab:sys_cfg}
\centering
\begin{tabular}{|c|l|}
    \hline
    \textbf{Host CPU} & AMD EPYC 9334, 32-Core, 2.7-3.9 GHz, \\
                      & 64 KB (per core) L1 cache, \\
                      & 1MB (per core) L2 cache, 128 MB shared L3 cache \\
    \hline
    \textbf{NDP} & DDR5-4800, 2 or 6 channels, 2 DIMMs per channel, \\                  & 2 ranks per DIMM, 2 VPEs and LNCs  per rank, \\
                & 256KB LNC-D, 8KB LNC-T, 1.2 GHz \\
    \hline
    \end{tabular}
    \vspace{-1em}
\end{table}
\end{scriptsize}

\subsubsection{Competing designs}
\begin{itemize}
    \item \textbf{CPU baselines:} HNSW and \textit{SCANN}~\cite{guo2020scann} on a 32-core CPU (\textit{CPU-baseline}) and a 96-core CPU (\textit{CPU-HP}).
    \item \textbf{Prior accelerator designs:} \textit{ANNA}~\cite{lee2022anna} on ASIC, \textit{DF-GAS}~\cite{zeng2023dfgas} on FPGA, \textit{PIMANN}~\cite{wu2025PIMANN} on UPMEM, and \textit{CAGRA}~\cite{ootomo2024cagra} on NVIDIA A100 GPU.
    \item \textbf{NDP baselines:} Vanilla HNSW on NDP (\textit{NDP-baseline}) and the SOTA NDP design \textit{ANSMET}~\cite{li2025ansmet}.
\end{itemize}

\begin{table}[t]
    \caption{\textbf{Specifications of Benchmark Datasets}.}
    \label{tab:dataset_features}
    \centering
    \resizebox{\linewidth}{!}{
    \begin{tabular}{|@{\hskip 3pt}c@{\hskip 3pt}|c@{\hskip 3pt}|c@{\hskip 3pt}|c@{\hskip 3pt}|c@{\hskip 3pt}|}
    \hline
        Dataset & Distance & \# Dims & \# Vectors & \# Queries  \\ \hline
        SIFT\cite{lowe2004sift} &  $\normltwo$ norm & 128 & 1M & 10K  \\ \hline
        GIST\cite{lowe2004sift} &  $\normltwo$ norm & 960 & 1M & 1K  \\ \hline
        BigANN\cite{bigann} &  $\normltwo$ norm & 128 & 1B & 10K    \\ \hline
        GloVe\cite{pennington2014glove} & IP & 100 & 1.2M & 1K  \\ \hline
        Wiki\cite{wiki} &  $\normltwo$ norm & 768 & 1M & 10K  \\ \hline
        MS\_MARCO\cite{Tri2016MSMACRO} & $\normltwo$ norm & 384 & 8M & 1K  \\ \hline
    \end{tabular}
    }
\end{table}

\subsubsection{Datasets}
The datasets used in this work are summarized in \cref{tab:dataset_features}.
SIFT, GIST, BigANN, and GloVe are standard ANNS datasets with high-dimensional vectors.
Wiki and MS\_MARCO are retrieval corpora.
Wiki contains Wikipedia articles, whose 768-dimensional embeddings are generated by Sentence-BERT \cite{reimers-2019-sentence-bert}.
MS\_MARCO consists of real Bing question--answer pairs, whose 384-dimensional embeddings are generated by the widely used BGE model \cite{bge_embedding}.
We build the indices of these datasets and convert the indices into the format required by \sysname.

\subsection{Overall Search Performance}
\label{eva:overall}

\sethlcolor{hl_pink}
\begin{figure}[t]
    \centering
    \includegraphics[width=\linewidth]{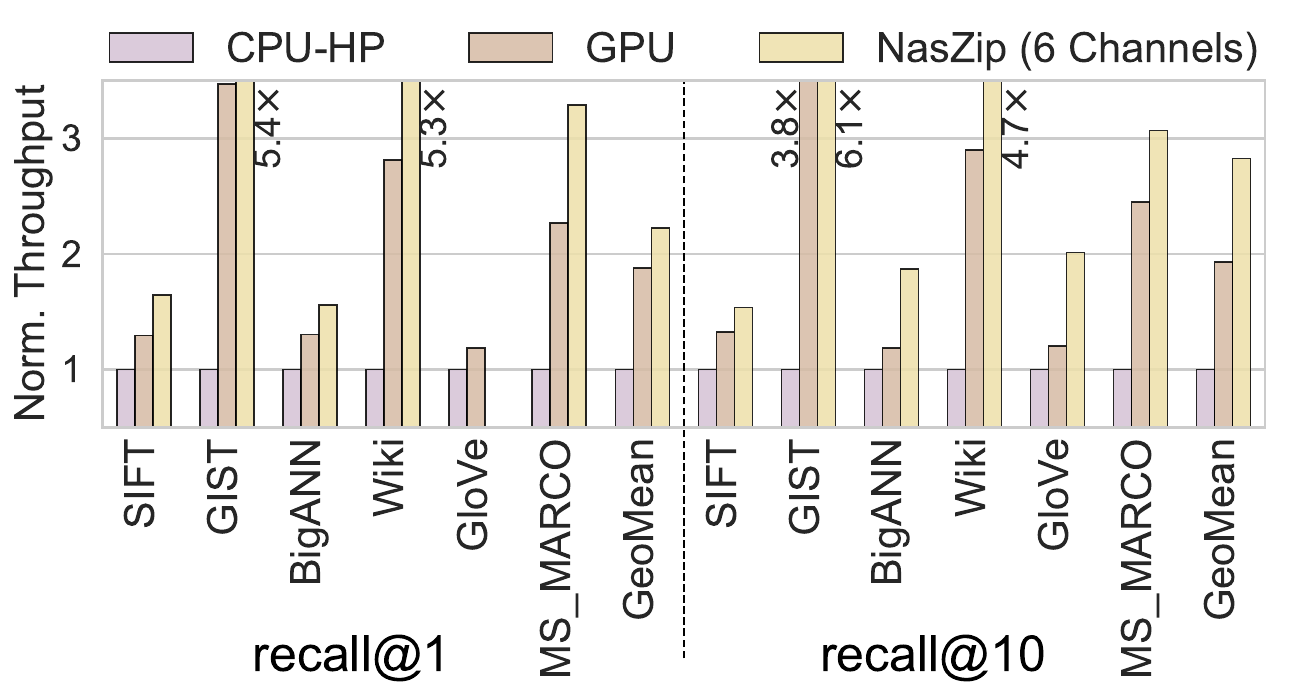}
    \caption{\textbf{Normalized throughput} (QPS) of CPU-HP, GPU and \sysname (6 channels), with recall@1 and recall@10$\geq90\%$.
    }
    \label{fig:eva_QPS_scale}
\end{figure}
\sethlcolor{yellow}

\subsubsection{Throughput}

\cref{fig:eva_overall_QPS} reports the speedup of \sysname with 2 channels (16 sub-channels) over comparable designs at a similar scale, with all results normalized to \design{CPU-Baseline}.
\cref{fig:eva_QPS_scale} compares \sysname with 6 channels against \design{GPU-Baseline} and \design{CPU-HP}.
\sysname is configured in \cref{fig:eva_QPS_scale} with 48 sub-channels, providing an aggregated memory bandwidth of 921.6~GB/s (19.2~GB/s per sub-channel).
The 48 sub-channels are organized as 6 channels $\times$ 2 DIMMs per channel $\times$ 2 ranks per DIMM $\times$ 2 sub-channels per rank.
\sethlcolor{yellow}

As shown in \cref{fig:eva_overall_QPS,fig:eva_QPS_scale}, \sysname consistently delivers the best throughput among prior ANNS designs across CPU, GPU, ASIC, UPMEM, FPGA, and NDP platforms.
It achieves an 8.4$\times$ speedup over the state-of-the-art CPU implementation \design{SCANN} and nearly 2$\times$ over the ASIC design \design{ANNA}.
Compared with the state-of-the-art NDP accelerator \design{ANSMET}, \sysname attains up to 1.69$\times$ higher performance through its tighter software--hardware co-design, particularly the more aggressive FEE-sPCA optimization.
It also outperforms \design{CPU-HP} and \design{GPU-Baseline} by 2.7$\times$ and 1.4$\times$, respectively, while substantially surpassing the UPMEM-based \design{PIMANN} despite PIMANN's high raw bandwidth.
The largest gain is observed on GIST.
This is because, as shown in \cref{fig:V_vs_Freq}, most of its early exits occur before dimension 193, pruning nearly 80\% of its 960 dimensions, whereas SIFT prunes only about 50\%.

\sethlcolor{hl_pink}
\begin{figure}[t]
    \centering
    \includegraphics[width=\linewidth]{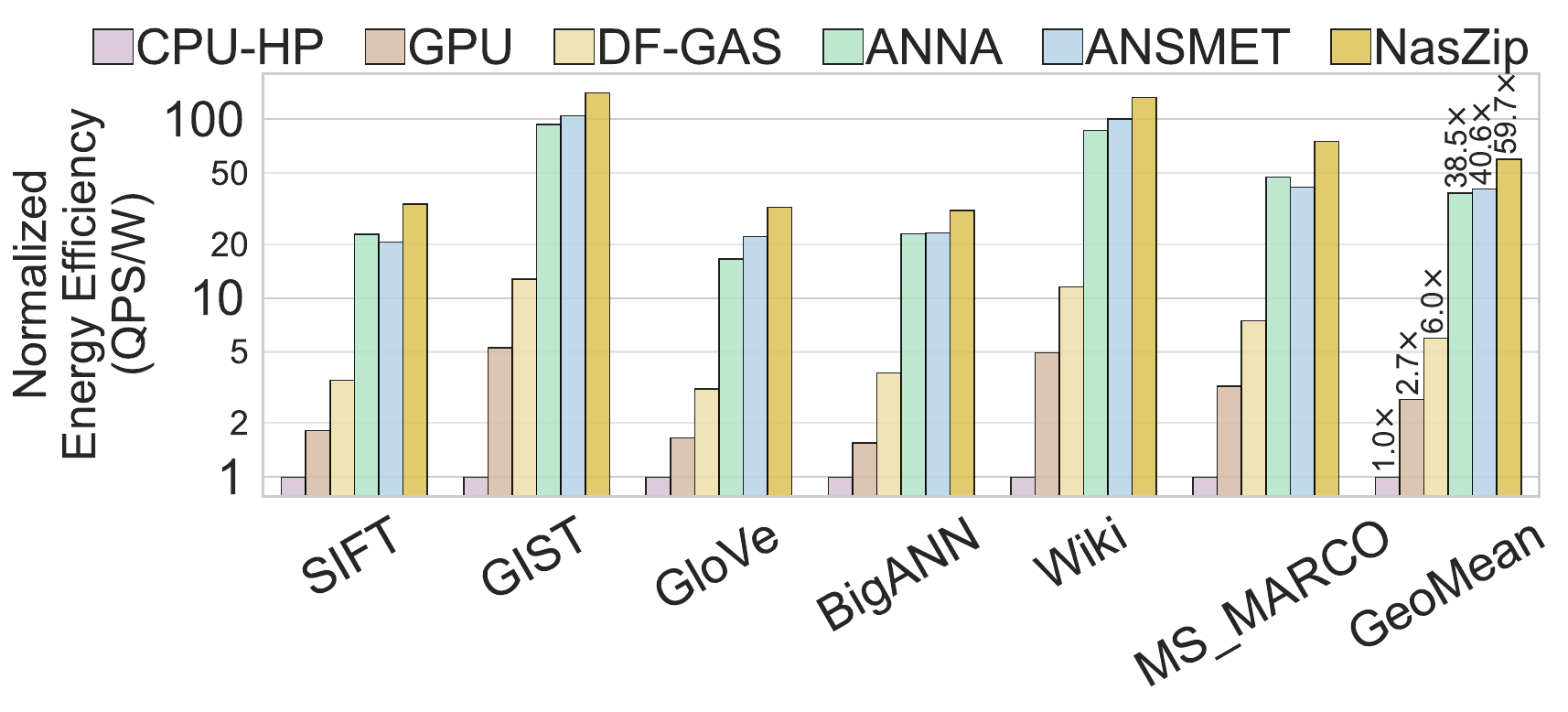}
    \caption{\textbf{Normalized energy efficiency} with recall@10$\geq90\%$.}
    \label{fig:eva_QPS_W}
\end{figure}
\sethlcolor{yellow}

\subsubsection{Energy efficiency}

The evaluation is shown in \cref{fig:eva_QPS_W}.
\design{GPU-Baseline} and \design{DF-GAS} achieve lower energy efficiency due to the high power consumption of HBM.
\design{ANNA} exhibits energy efficiency comparable to that of the NDP design \design{ANSMET}.
By enabling more aggressive early exiting (FEE-sPCA), reducing cross-channel communication (DaM) and caching of frequently accessed neighbor lists (LNC), \sysname achieves up to 1.5$\times$ higher energy efficiency than \design{ANSMET}.

\sethlcolor{hl_pink}
\begin{figure}[t]
    \centering
    \includegraphics[width=\linewidth]{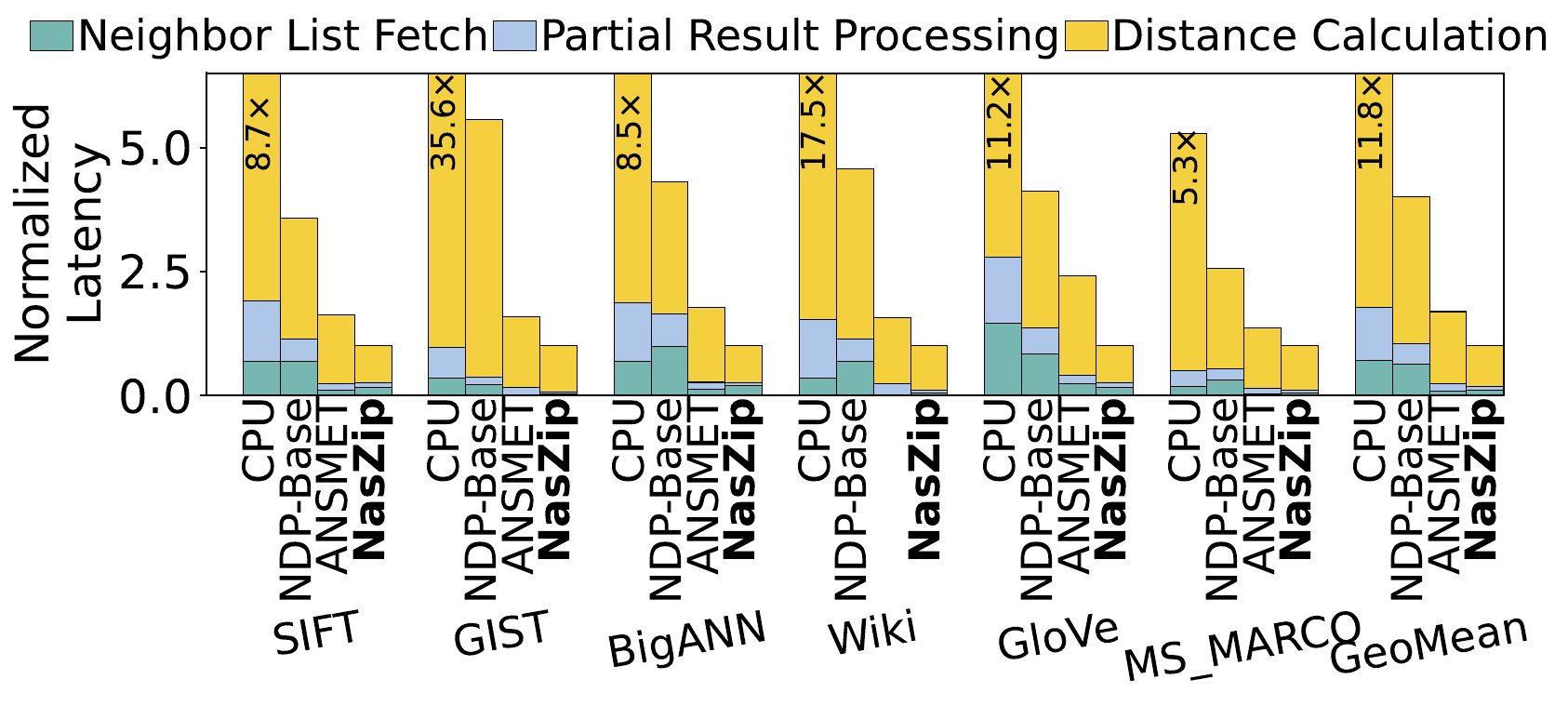}
    \vspace{-1.5em}
    \caption{\textbf{Latency comparison and breakdown} (normalized to \sysname) with recall@10$\geq90\%$.}
    \vspace{-1em}
    \label{fig:eva_latency}
\end{figure}
\sethlcolor{yellow}

\subsection{In-depth Analysis}
\label{sec:eva_detailed}

\subsubsection{Latency Breakdown}

\cref{fig:eva_latency} breaks down query latency into neighbor-list retrieval, distance computation, and partial-result processing (including CPU-NDP communication in NDPs).
FEE-sPCA and Dfloat substantially reduce distance-computation latency, while the local neighbor cache keeps hot indices on NDP, accelerating neighbor-list retrieval and further reducing CPU-NDP communication overhead.

\subsubsection{Throughput versus Recall}
\begin{figure}[t]
    \centering
    \includegraphics[width=\linewidth]{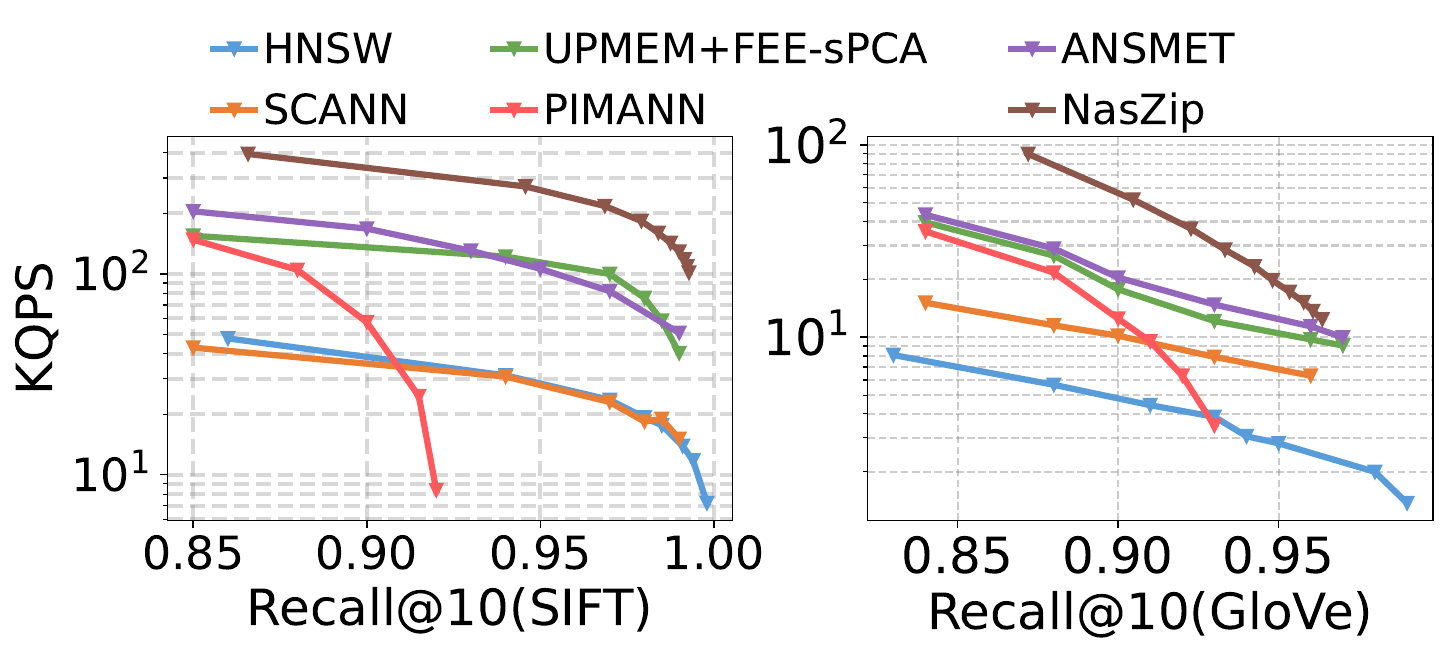}
    \caption{\textbf{Comparison of throughput versus recall.}}
    \label{fig:eva_overall_thp_recall}
\end{figure}

\cref{fig:eva_overall_thp_recall} evaluates the effect of varying the search range ($efSearch$).
Increasing $efSearch$ expands the search scope, improving recall but reducing QPS.
Overall, \sysname consistently outperforms baselines.

\subsubsection{Memory Traffic of Database Compression}

\sethlcolor{hl_pink}
\begin{figure}[t]
    \centering
    \includegraphics[width=\linewidth]{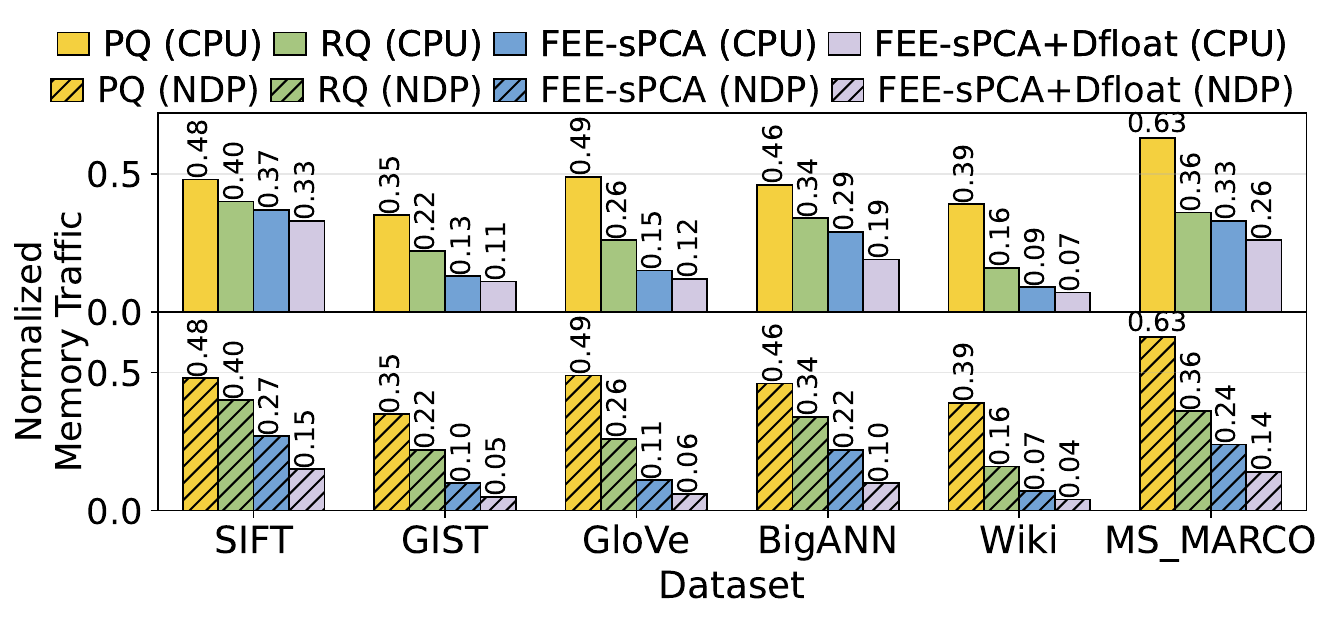}
    \caption{\textbf{Memory traffic comparison of database compression methods} (PQ and RabitQ on HNSW), evaluated with recall@10$\geq90\%$. Results are normalized to HNSW.}
    \label{fig:eva_compression_memory}
\end{figure}
\sethlcolor{yellow}

\cref{fig:eva_compression_memory} compares memory traffic against representative ANNS compression baselines on HNSW at recall@10$\geq 90\%$.
PQ \cite{pq2011} is mainly designed for compression and incurs substantial precision loss.
To maintain high recall, PQ must use a weaker compression ratio, leading to much higher memory traffic (about $2\times$ that of RabitQ and \sysname).
RabitQ~\cite{gao2024rabitq} accelerates candidate filtering with compact quantized vector representations, but surviving candidates still require exact full-dimensional distance computation during re-ranking.
In contrast, FEE-sPCA reduces memory traffic through feature-level early exiting, thereby cutting the number of accessed dimensions, while Dfloat further reduces the bit width of each accessed feature.
Meanwhile, FEE-sPCA and Dfloat are compatible with the memory access patterns on NDP.
As a result, our method achieves lower memory traffic at the same recall level, especially on NDP.

\begin{figure}[t]
    \centering
    \includegraphics[width=\linewidth]{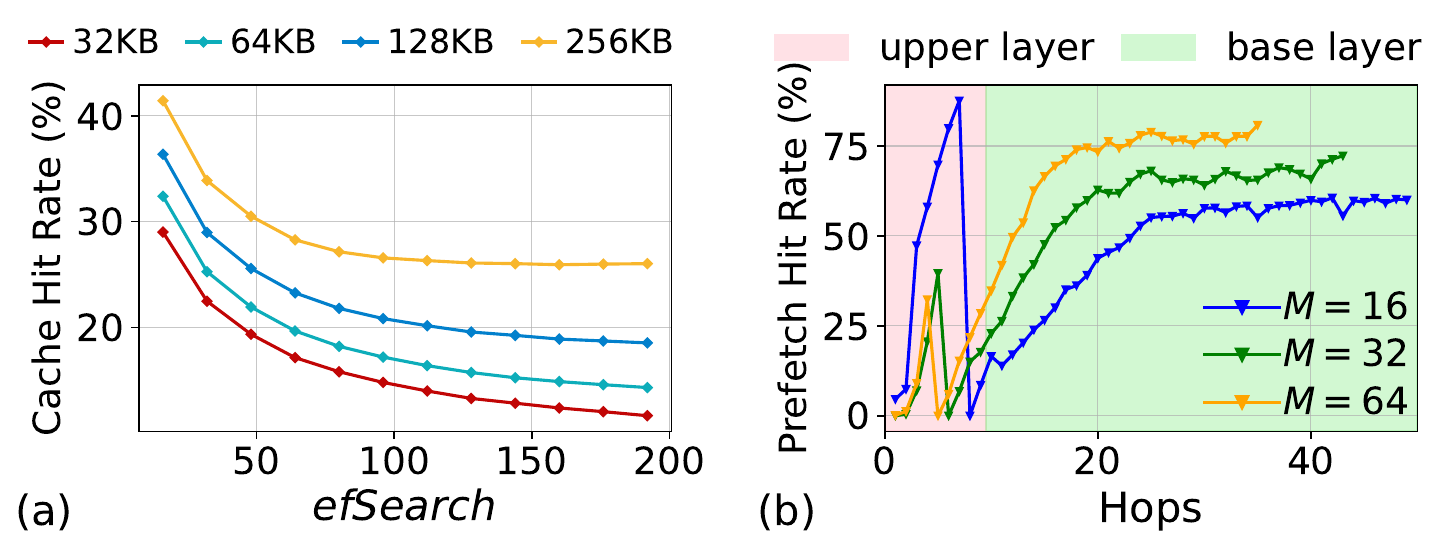}
    \caption{
    \textbf{(a) Hit rate of LNC-D versus search parameters} $efSearch$ on SIFT, with varying cache size.
    \textbf{(b) Average prefetch hit rate \wrt\ search hops}. Evaluated using 1M queries with different graph construction parameters $M$.}
    \label{fig:eva_detail_4}
\end{figure}

\subsubsection{Cache Size of LNC}
\label{subsec:lnc_size}

\cref{fig:eva_detail_4}a evaluates the impact of LNC-D capacity on cache hit rate.
\sysname adopts a 256KB LNC-D, and we vary the enabled capacity to analyze its impact.
Overall, larger LNC-D capacity leads to a higher hit rate by retaining more frequently accessed neighbor lists.
As $efSearch$ increases, the hit rate decreases because a larger search range visits more diverse candidate nodes and weakens temporal locality.
Beyond a certain point ($efSearch > 50$), most hot neighbor lists are already retained in LNC-D, and the additional cache misses mainly come from a small number of low-reuse tail nodes, causing the hit rate to converge.

\subsubsection{Prefetching performance}
\label{subsubsec:eva_prefetch}
\cref{fig:eva_detail_4}b profiles the prefetch hit rate at each hop and its dependence on graph density, controlled by $M$.
The hit rate gradually increases in the upper layers but drops when entering the base layer, because upper-layer neighbor lists differ from base-layer ones and thus invalidate cached entries.
As $M$ increases, the hit rate decreases in the upper layers but rises in the base layer: a wider upper-layer search identifies most nearest neighbors earlier, stabilizing the candidate queue and reducing updates in the base layer.
Overall, the prefetch hit rate remains above 50\%.

\subsubsection{Performance versus Batch Size}
\label{subsec:batch_eva}

\begin{figure}[t]
    \centering
    \includegraphics[width=\linewidth]{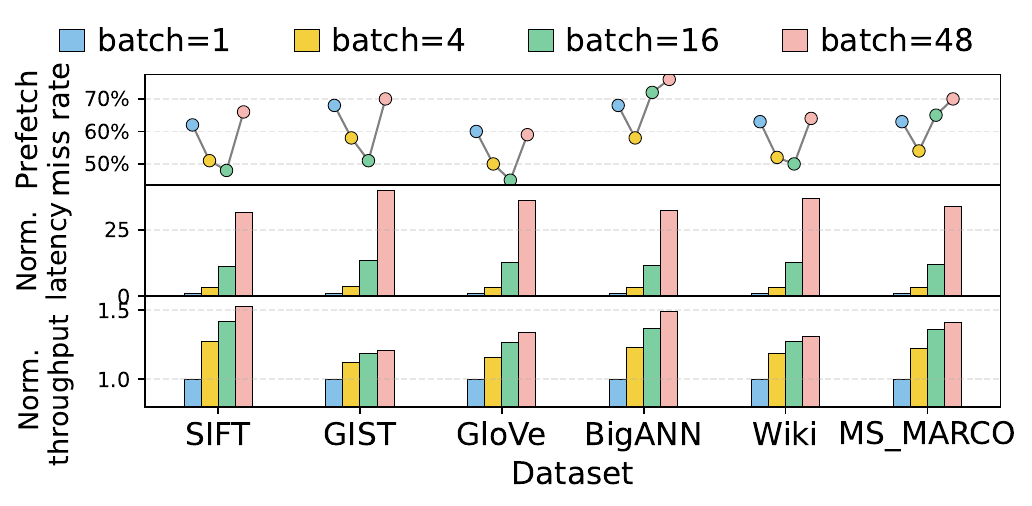}
    \caption{\textbf{Prefetch miss rate, latency and throughput versus batch sizes}, evaluated under recall@10$\geq90\%$.}
\label{fig:eva_batch}
\end{figure}

\cref{fig:eva_batch} evaluates throughput, latency and relative prefetch miss rate under different batch sizes.
As batch size increases, throughput improves due to better sub-channel utilization and higher cache reuse.
However, latency also increases, especially when the batch size grows from 16 to 48.
This is because prefetching is most effective at batch size 16, whereas at batch size 48, excessive prefetch misses increase cache contention and reduce its benefit.
To balance throughput and latency, we use batch size 16 in all other evaluations.

\begin{figure}[t]
    \centering
    \includegraphics[width=\linewidth]{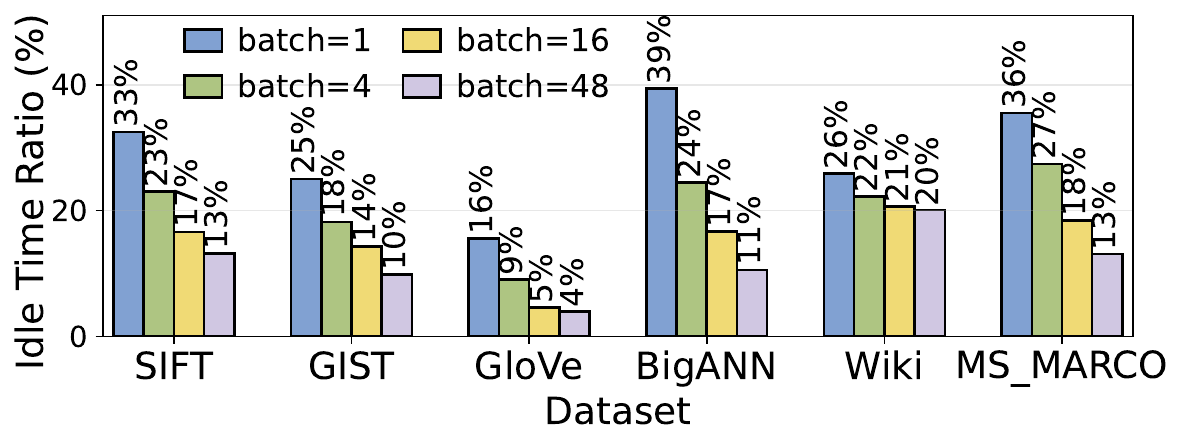}
    \caption{\textbf{Idle time ratio of the earliest finishing sub-channel.}}
\label{fig:eva_workload_balance}
\end{figure}

\subsubsection{Workload Balance Analysis}
\label{subsec:loadbalance}

\cref{fig:eva_workload_balance} reports the average idle time of the least-loaded sub-channel (\ie, the earliest finishing one), normalized to total execution time.
The workload imbalance is more severe at small batch sizes, for example, on BigANN, the idle time reaches 39\% when the batch size is 1.
As the batch size increases, the imbalance decreases, since larger batches average out the variation in the total number of vector dimensions processed by different sub-channels.
However, Wiki shows higher imbalance than the other datasets.
This is because the other datasets are shuffled to improve distribution uniformity, whereas Wiki is left unshuffled to preserve the spatial and semantic locality of consecutive document chunks for better retrieval quality, consistent with practical RAG deployments~\cite{sarthi2024raptor}.
As a result, Wiki accesses are more clustered across sub-channels, leading to higher workload imbalance.

\subsection{End-to-end RAG Evaluation}
\label{subsec:eva_RAG}

\begin{figure}[t]
    \centering
    \includegraphics[width=\linewidth]{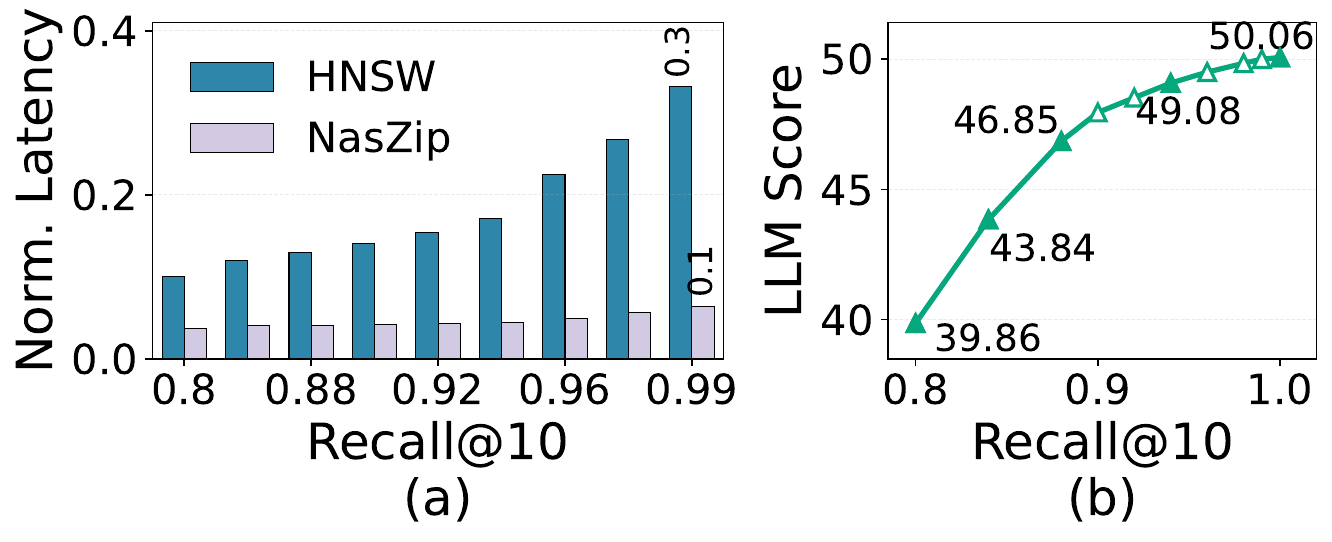}
    \caption{(a) \textbf{LLM latency (TTFT) \vs retrieval accuracy (Recall@10)}, normalized to KNN baseline. (b) \textbf{LLM answer quality (RAGAS score) \vs retrieval accuracy (Recall@10)}.}
    \label{fig:eva_end_to_end}
\end{figure}

\cref{fig:eva_end_to_end} evaluates the RAG end-to-end using GPT-4o.
The corpora are drawn from 2WikiMultihopQA \cite{xanh2020_2wikimultihop}, HotpotQA \cite{yang2018hotpotqa}, MultiFieldQA-en \cite{Multifieldqa}, QASPER \cite{Dasigi2021ADO}, and MS\_MARCO \cite{Tri2016MSMACRO}.
To preserve retrieval quality, we use the text-embedding-ada-002 \cite{openai2022embedding} model from OpenAI, which produces 1536-dimensional embeddings.
\cref{fig:eva_end_to_end}a shows latency (time-to-first-token, TTFT) versus recall@10, using KNN search as the baseline.
\sysname substantially reduces latency and retains significant speedup even under high-recall requirements.
\cref{fig:eva_end_to_end}b shows RAG quality under different retrieval accuracy levels (recall@10).
Quality is measured by the LLM score from RAGAS \cite{ragas2024}, reflecting answer correctness and hallucination.
When recall@10 exceeds 0.9, response quality degrades only marginally \wrt\ the ideal case of recall@10=1.
Overall, \sysname is robust enough to maintain high RAG quality while significantly reducing latency.

\begin{figure}[t]
    \centering
    \includegraphics[width=\linewidth]{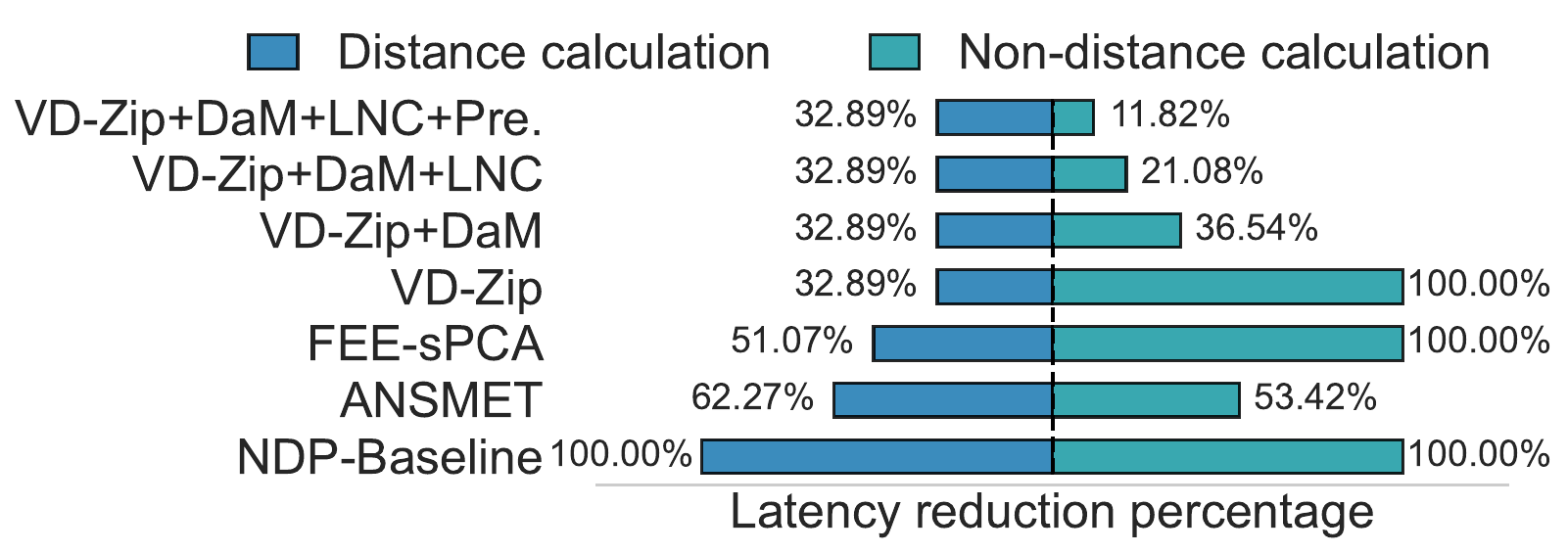}
    \caption{
    \textbf{Latency reduction from each \sysname optimization}, compared with ANSMET. From bottom to top, each represents the latency reduction compared to the baseline.
    }
    \label{fig:eva_ablation_breakdown}
\end{figure}

\subsection{Ablation study}
\label{sec:ablation_study}

\cref{fig:eva_ablation_breakdown} shows how each technique reduces both distance and non-distance latency.
For reference, ANSMET reduces distance calculation latency to 62.27\% through bit-level FEE and non-distance latency to 53.42\% through mapping and scheduling.
\sysname further reduces distance latency to 51.07\% with FEE-sPCA, while Dfloat provides an additional 1.79$\times$ speedup.
For non-distance overheads, DaM and LNC-T/D reduce latency to 36.54\% and 21.08\%, respectively, and prefetching (Pre.) further cuts it by about 50\%.
This highlights the effectiveness of neighbor-list caching and prefetching.

\begin{table}[t]
\belowrulesep=0pt
\aboverulesep=0pt
\caption{\textbf{Offline and online overhead of PCA-based pre-processing for database and query.}}
\label{tab:pca_overhead}
\footnotesize
\centering
\setlength{\tabcolsep}{3.5pt}
\renewcommand{\arraystretch}{1.08}
\begin{tabular}{lcccc}
Dataset & \makecell{Size /\\ Dim.} & \makecell{Offline\\ time (s)} & \makecell{Online\\ latency (ms)} & \makecell{Online\\ overhead (\%)} \\
\midrule[1pt]
SIFT      & 1M / 128   & 6.54   & 0.149 & 3.6 \\
GIST      & 1M / 960    & 53.27  & 0.817 & 0.4 \\
BigANN    & 1B / 128   & 430.66 & 0.135 & 1.7 \\
GloVe     & 1.2M / 100 & 5.23   & 0.127 & 0.1 \\
MS\_MARCO & 8M / 384   & 30.91  & 0.519 & 3.8 \\
Wiki      & 1M / 768   & 40.94  & 0.727 & 3.2 \\
\bottomrule
\end{tabular}
\end{table}

\subsection{Overhead Analysis}

\subsubsection{PCA Preprocessing}

During the offline phase, FEE-sPCA requires database preprocessing, mainly to compute PCA eigenvalues, which introduces additional overhead.
\cref{tab:pca_overhead} reports the preprocessing time on an A100 GPU.
Although the overhead increases with dataset size, it typically remains on the order of seconds to minutes and is small compared with index construction time (\eg, building HNSW on BigANN takes hours).
During the online phase, queries must also be PCA-transformed at the embedding stage.
As shown in \cref{tab:pca_overhead}, this one-shot transformation adds negligible overhead \wrt\ the entire search latency.
\sethlcolor{yellow}

\begin{figure}[t]
\begin{subfigure}[h]{0.47\linewidth}
\centering
\resizebox{\columnwidth}{!}{
\fontsize{17}{20}\selectfont
\begin{tabular}{ l | r }
\hline
\textbf{\makecell{Component}} &  \textbf{\makecell{Area($\bm{\mu}$m\textsuperscript{2})}} \\
\hline
\hline
{\sysname Add-on}                  & {} \\
{\quad$\triangleright$ LNC-D} & {489.6K}  \\
{\quad$\triangleright$ LNC-T}  & {37.5K}  \\
{\quad$\triangleright$ VPE}  & {144.6K} \\
{\quad$\triangleright$ Controller}  & {9.9K} \\
{\quad$\triangleright$ Query Buffer}  & {17.1K} \\
{\quad$\triangleright$ Others}     & {10.4K}  \\
\hline
Total & 709.1K \\
\hline
\end{tabular}
}
    \label{tab:area_breakdown}
\end{subfigure}
    \hfill
\begin{subfigure}[h]{0.5\linewidth}
    \centering
    \includegraphics[width=\linewidth]{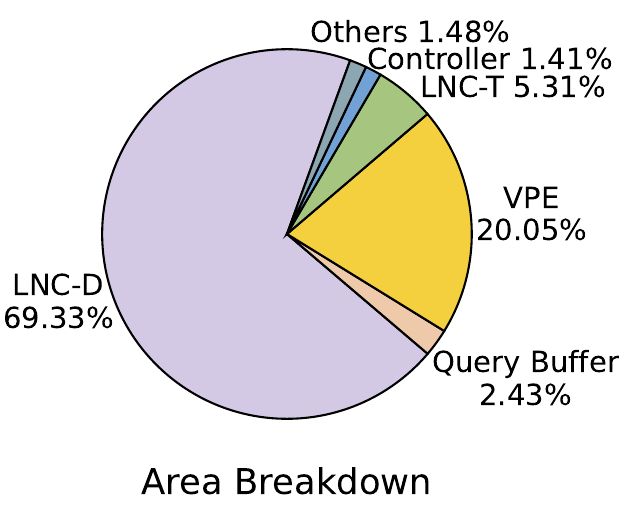}
\end{subfigure}
    \caption{\textbf{Area overhead of added components in \sysname.}}
    \label{fig:area_breakdown}
\end{figure}

\begin{figure}[t]
    \centering
    \includegraphics[width=\linewidth]{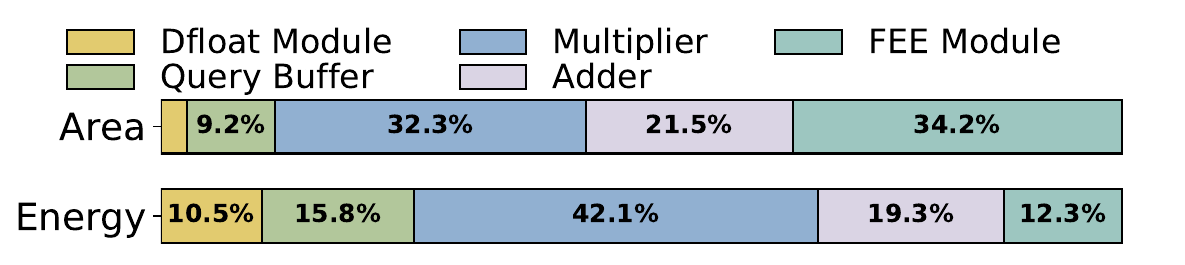}
    \caption{\textbf{Area and energy breakdown of VPE modules.}}
    \label{fig:vdp_overehad}
\end{figure}

\subsubsection{Area and Energy overhead}
\label{eva:area_power_overhead}

The area overhead of the additional NDP components in each sub-channel is shown in \cref{fig:area_breakdown}.
The total area overhead of \sysname is 0.7091~mm\textsuperscript{2}, which is marginal compared with the 10.22~mm\textsuperscript{2} area of the standard RCD \cite{rcd} and DB \cite{db} components.
\cref{fig:vdp_overehad} further breaks down the VPE overhead introduced by FEE-sPCA and Dfloat.
The \textit{Query Buffer} and \textit{FEE Module} dominate the area due to query and parameter storage, while the \textit{Multiplier} and \textit{Adder} dominate energy consumption because they remain active for most of the execution.

\subsubsection{Thermal Impact}
We further use 3D-ICE \cite{zhu2026threeice} to evaluate the thermal impact of our design.
At an ambient temperature of 28$^\circ$C, the combined heat from the added logic and DRAM results in a peak DRAM-cell temperature of 65.47$^\circ$C.
According to JEDEC specifications \cite{jedec_ddr5b}, the default refresh mode provides sufficient data retention for the standard refresh interval ($t_\textrm{REFI}$) at temperatures up to 85$^\circ$C.
Therefore, \sysname~does not compromise DRAM reliability, even without active cooling.

\section{Related Work}

\textbf{Software-based ANNS Acceleration.}
Early research focused on tree-based (\eg, R-tree \cite{guttman1984Rtree},
KD-tree \cite{fried1977kdtree}) and hash-based (\eg, LSH
\cite{datar2004LSH,gan2012C2LSH}) approaches
to optimize ANNS index structure.
Subsequently, quantization-based methods (\eg, PQ \cite{pq2011,ge2014opq}, RabitQ \cite{gao2024rabitq}) were proposed to reduce the index size and calculation overhead
by pre-computing some values during the index-building stage.
In addition, graph-based methods (\eg, NSG \cite{fu2019nsg}, HNSW \cite{malkov2018hnsw}) are widely adopted due to their higher accuracy and speed.
Recent works also propose dimension reduction
\cite{gao2023adsampling,gao2024rabitq} and reordering \cite{wang2024leanor}
methods.
Meanwhile, works such as SCANN \cite{guo2020scann}, SPFresh \cite{xu2023spfresh}, and VBASE \cite{zhang2023vbase} integrate optimized indexing, quantization, updating and query-processing techniques to achieve high performance on CPU.

\textbf{Hardware-based ANNS Acceleration.}
CAGRA \cite{ootomo2024cagra} optimizes graph-based ANNS on GPU, achieving up to one million QPS.
ANNA \cite{lee2022anna} and NeuVSA \cite{Yuan2025NeuVSA} are ASIC designs targeting the quantization-based ANNS (PQ).
DF-GAS \cite{zeng2023dfgas} proposes accelerating graph-based ANNS on FPGA, achieving high throughput by exploring feature-packing memory access patterns and a parallel search scheme.
DiskANN~\cite{diskann}, SPANN~\cite{chen2021spann}, and SPFresh~\cite{xu2023spfresh} leverage SSD-backed or disk-based indices to support billion-scale vector search with reduced DRAM requirements.
Some designs are implemented based on near-SSD computation including VStore
\cite{vstore2022}, NDSearch \cite{wang2024ndsearch}, SmartANNS
\cite{Bing2024SmartSSDs}, REIS \cite{reis2025}, and ICE \cite{hu2023ice}.
They achieve better results than disk designs, but the SSD speed is still slower than DRAM.
Designs based on near/in-data processing emerge, as they provide promising bandwidth.
Some works like UPVSS \cite{liu2025UPVSS} and PIMANN \cite{wu2025PIMANN} implement ANNS acceleration with the help of UPMEM PIM, while DRIM-ANN \cite{DRIM-ANN} targets commercial DRAM-PIM.
ANSMET \cite{li2025ansmet}, CXL-ANNS \cite{Junhyeok2023CXL-ANNS} and DReX \cite{quinn2025drex} further employ near/in-data processing and hardware/software co-designs to accelerate ANNS and dense retrieval.
ANSMET employs DIMM-based NDP and implements hybrid early exiting.
However, its early exiting threshold is not sufficiently strict, which limits its performance.
\sysname further boosts performance by using FEE-sPCA and Dfloat to eliminate more redundant computations, while leveraging the combined hardware optimizations of DaM and LNC.

\section{Conclusion}

Graph-based ANNS is widely adopted in vector databases for its high accuracy and low latency, but its memory-bound nature makes memory bandwidth critical to performance.
\sysname addresses this challenge through an efficient NDP architecture and a software-hardware co-design for ANNS acceleration.
Our software innovations include statistics-based early exiting and dynamic floating-point representation.
Our hardware innovations include data-aware mapping, caching, and prefetching.
Together, they significantly improve the performance over baselines.
Consequently, \sysname outperforms state-of-the-art ANNS designs across diverse architectures.

\appendix
\section{Artifact Appendix}

\subsection{Abstract}

The evaluation contains two major parts: the overall performance of \sysname under two different configurations and the detailed analysis of QPS \vs recall.

We conduct our evaluation on several datasets, including SIFT, GIST, BigANN, GloVe, Wiki, and MS\_MARCO.
We provide the source code for the proposed algorithms (FEE-sPCA and Dfloat), as well as the hardware simulator (UniNDP) and all corresponding configuration files.
The initial HNSW indexes are built using NVIDIA’s cuVS library to ensure high index quality.
However, index construction is time-consuming, especially for BigANN, and indexes generated by different versions of cuVS on different GPUs may introduce variations in the results.
To facilitate fast reproduction, we therefore provide pre-built indexes.
At the same time, we also provide the code and instructions for building the indexes from scratch.
Our evaluation is conducted on a CPU server with 256 GB of memory. For faster reproduction, more CPU cores and larger memory capacity are preferred.
If the indexes are built from scratch, a GPU with more than 24 GB of memory is required for all datasets except BigANN, while more than 70 GB is required for BigANN.

\subsection{Artifact check-list (meta-information)}

{\small
\begin{itemize}
  \item {\bf Data set:} SIFT, GIST, BigANN, GloVe, Wiki, MS\_MARCO.
  \item {\bf Run-time environment}: Ubuntu 22.04 LTS, CUDA 12.x (required only when building the index from scratch)
  \item {\bf Hardware:} A server with an x86 processor and at least 128\,GB of DRAM.
  Building indexes from scratch additionally requires an NVIDIA GPU with at least 24\,GB of VRAM for datasets other than BigANN. For BigANN, index construction from scratch requires at least 70\,GB of VRAM and 320\,GB of DRAM.
  \item {\bf Metrics:} recall, QPS.
  \item {\bf Output:} recall and QPS of \sysname across several datasets.
  \item {\bf How much disk space required (approximately):}
  If using the pre-built indexes, 150 GB of storage is required. If building the indexes from scratch, 200 GB of storage is required.
  \item {\bf How much time is needed to prepare workflow (approximately):} It needs about 1 hour to download pre-built indexes (About 85 GB).
  \item {\bf How much time is needed to complete experiments (approximately):} It needs about 7 hours (16 cores parallel simulation) with pre-built indexes.
  If building the indexes from scratch, about another 1 hour is required.
  \item {\bf Publicly available: } It is publicly available on GitHub \url{https://github.com/Intelligent-Computing-Research-Group/NasZip}
  \item {\bf Code licenses:} Apache-2.0 license.
  \item {\bf Data licenses:} The datasets are publicly available through their original licensing terms.
  \item {\bf Archived (provide DOI): } \url{https://doi.org/10.5281/zenodo.19453078}
\end{itemize}
}

\subsection{Description}

\subsubsection{How to access}
We archive the source code at \url{https://doi.org/10.5281/zenodo.19453078}.
We recommend accessing our GitHub repository: \url{https://github.com/Intelligent-Computing-Research-Group/NasZip} for the latest version.

\subsubsection{Hardware dependencies}
If using the pre-built indexes, the minimum hardware requirement is a server with an x86 processor, at least 16 CPU cores, and at least 128\,GB of DRAM.
If building indexes from scratch, an additional NVIDIA GPU with at least 24\,GB of VRAM is required for datasets except BigANN. For BigANN, building the index from scratch requires at least 70\,GB of VRAM and 320\,GB of DRAM.

\subsubsection{Software dependencies}
If using the pre-built indexes and evaluating on a CPU server:
\begin{itemize}
    \item Ubuntu 22.04
    \item Conda 25.9.1
    \item Python 3.12
    \item PyTorch 2.7.1+cpu
\end{itemize}

If building the indexes from scratch on a GPU server:
\begin{itemize}
    \item Ubuntu 22.04
    \item Anaconda 24.4.0
    \item Python 3.12
    \item PyTorch 2.5.1+cu124
    \item CuPy 12.3.0
\end{itemize}

\subsubsection{Data sets}
SIFT, GIST, BigANN, and GloVe are standard ANNS datasets with high-dimensional vectors.
Wiki and MS\_MARCO are retrieval corpora.
Wiki contains Wikipedia articles, whose 768-dimensional embeddings are generated by Sentence-BERT \cite{reimers-2019-sentence-bert}.
MS\_MARCO consists of real Bing question--answer pairs, whose 384-dimensional embeddings are generated by the widely used BGE model \cite{bge_embedding}.

\subsection{Installation}

We provide a well-documented README file with detailed installation instructions.
Specifically, users are guided to first create a virtual environment, then install the required packages and dependencies, and finally download the pre-built indexes.

\subsection{Evaluation and expected results}
We reproduced the key results of \sysname, specifically those presented in \cref{fig:eva_overall_QPS}, \cref{fig:eva_QPS_scale}, \cref{fig:eva_latency}, \cref{fig:eva_overall_thp_recall} and \cref{fig:eva_detail_4}.

\subsection{Notes}
We recommend using the pre-built indexes for simulation, as they can reproduce the same results reported in the paper.
If the indexes are built from scratch, the graph construction in cuVS involves randomness, which may lead to slight variations in the results.

\subsection{Methodology}

Submission, reviewing and badging methodology:

\begin{itemize}
  \item \url{https://www.acm.org/publications/policies/artifact-review-and-badging-current}
  \item \url{https://cTuning.org/ae}
\end{itemize}

% Finally, the ACM/IEEE Plagiarism Policies\footnote{\url{http://www.acm.org/publications/policies/plagiarism_policy}\\
% \url{https://www.ieee.org/publications_standards/publications/rights/plagiarism_FAQ.html}}
% cover a range of ethical issues concerning the misrepresentation of
% other works or one's own work.

% \section*{Acknowledgements}
% This document is derived from previous conferences, in particular ISCA 2019,
% MICRO 2019, ISCA 2020, MICRO 2020, ISCA 2022, HPCA 2022, and ISCA 2024. 

%%%%%%% -- PAPER CONTENT ENDS -- %%%%%%%%

%%%%%%%%% -- BIB STYLE AND FILE -- %%%%%%%%
% \bibliographystyle{IEEEtranS}
\bibliographystyle{IEEEtran}
\bibliography{refs}
%%%%%%%%%%%%%%%%%%%%%%%%%%%%%%%%%%%%

\end{document}